\documentclass[11pt,a4paper,fleqn]{article}
\pdfoutput=1   % required for arXiv submission if pdflatex is used
\usepackage[utf8]{inputenc}
\usepackage{lmodern}
\usepackage{alpha}
\usepackage{parskip}
\usepackage{enumitem}
\usepackage{xcolor}
\usepackage{extarrows}
\usepackage{bm}

\renewcommand{\vec}[1]{\mathbf{#1}}
\DeclareMathOperator{\supp}{supp}

\renewcommand{\sin}{{\mathrm{in}}}
\newcommand{\sout}{{\mathrm{out}}}
\newcommand{\as}{{\mathrm{as}}}
\newcommand{\rvac}{| 0 \rangle}
\newcommand{\lvac}{\langle 0 |}

\makeatletter
\newcommand{\superimpose}[2]{{%
  \ooalign{%
    \hfil$\m@th#1\@firstoftwo#2$\hfil\cr
    \hfil$\m@th#1\@secondoftwo#2$\hfil\cr
  }%
}}
\makeatother

\newcommand{\rprod}{\operatornamewithlimits{ \mathbin{\mathpalette\superimpose{{\hspace{0.3em}\xlongrightarrow{\hphantom{\hspace{0.4em}\prod}}\vphantom{\prod}}{\prod}} } }}
\newcommand{\lprod}{\operatornamewithlimits{ \mathbin{\mathpalette\superimpose{{\hspace{0.2em}\xlongleftarrow{\hphantom{\hspace{0.4em}\prod}}\vphantom{\prod}}{\prod}} } }}

\AtBeginDocument{
    \setlength{\jot}{8pt}
    \setlength{\abovedisplayskip}{12pt}
    \setlength{\belowdisplayskip}{12pt}
    \setlength{\abovedisplayshortskip}{12pt}
    \setlength{\belowdisplayshortskip}{12pt}
}

\begin{document}

% !TEX root = paper.tex

% \preprintno{%
% HU-EP-22/29-RTG\\
% \vfill
% }

\title{%
Scattering Amplitudes from Euclidean Correlators:\\
Haag-Ruelle theory and approximation formulae
}

\author[hu,desy]{Agostino Patella}
\author[TVrome,TVinfn]{Nazario Tantalo}

\address[hu]{Humboldt Universit\"at zu Berlin, Institut f\"ur Physik \& IRIS Adlershof, \\Zum Grossen Windkanal 6, 12489 Berlin, Germany}
\address[TVrome]{Universit\`a di Roma Tor Vergata, Dipartimento di Fisica, \\Via della Ricerca Scientifica 1, 00133 Rome, Italy}
\address[TVinfn]{INFN, Sezione di Tor Vergata, Via della Ricerca Scientifica 1, 00133 Rome, Italy}
\address[desy]{DESY, Platanenallee 6, D-15738 Zeuthen, Germany}

\begin{abstract}
In this work we provide a non-perturbative solution to the theoretical problem of extracting scattering amplitudes from Euclidean correlators in infinite volume. We work within the solid axiomatic framework of the Haag-Ruelle scattering theory and derive formulae which can be used to approximate scattering amplitudes arbitrarily well in terms of linear combinations of Euclidean correlators at discrete time separations. Our result generalizes and extends the range of applicability of a result previously obtained by Barata and Fredenhagen~\cite{Barata:1990rn}. We provide a concrete procedure to construct such approximations, making our formulae ready to be used in numerical calculations of non-perturbative QCD scattering amplitudes. A detailed numerical investigation is needed to assess whether the proposed strategy can lead to the calculation of scattering amplitudes with phenomenologically satisfactory precision with presently available lattice QCD data. This will be the subject of future work. Nevertheless, the numerical accuracy and precision of lattice simulations is systematically improvable, and we have little doubts that our approach will become useful in the future.

\end{abstract}

\begin{keyword}
    Quantum Field Theory,
    Non-perturbative $S$-matrix
\end{keyword}

\maketitle
% bold math in section headings
% (AFTER \maketitle, otherwise issue with affiliation counters until fixed)
\makeatletter
\g@addto@macro\bfseries{\boldmath}
\makeatother

\tableofcontents
\pagebreak

%--------------------------------------------------------------
\section{Introduction}\label{sec:intro}
%To spell-check, run:
%aspell --mode=tex --home-dir=. --conf=./.aspell.texconf --lang=en_US -x -c 00-introduction.tex

Scattering matrix ($S$-matrix) elements are the core observables of Quantum Field Theories (QFT) admitting particle interpretation. In the case of strongly interacting theories and, therefore, in the phenomenologically relevant case of Quantum Chromodynamics (QCD), scattering amplitudes cannot be calculated by using perturbative techniques. The non-perturbative accuracy required for their evaluation can, at least in principle, be obtained by means of lattice simulations. These are performed in Euclidean time, by introducing a finite volume and by discretizing it in units of the so-called lattice spacing. The lattice spacing and the finite volume regularize the theory in the ultraviolet and in the infrared. The Euclidean signature allows a probabilistic interpretation of the quantum mechanical path-integral of the theory and, hence, the numerical calculation of time-ordered Euclidean correlators by means of Markov-Chain Monte Carlo techniques. The Euclidean correlators contain all the physical information of a QFT. However, in practice, one needs to understand how to extract this information from a finite set of data, obtained on a finite volume and affected by numerical and statistical noise.

Physical quantities associated with stable single-particle states can be easily extracted from Euclidean correlators. This can be done by studying the leading asymptotic behavior at large Euclidean times of lattice correlators that, indeed, are dominated by single-particle contributions. Conversely, $S$-matrix elements involving multi-particle states have to be extracted from contributions to Euclidean correlators that are exponentially suppressed in Euclidean time~\cite{Maiani:1990ca}, and this poses challenging theoretical and numerical problems. Moreover, the continuum part of the spectrum gets quantized in a finite volume. In this setup, energy eigenstates heavier than single-particle states cannot be interpreted as states of incoming or outgoing asymptotic particles.

In a series of ground-breaking papers~\cite{Luscher:1991cf,Luscher:1990ux,Luscher:1990ck,Luscher:1985dn}, L\"uscher managed to turn the infrared problem of the quantization of the spectrum into a non-perturbative method to compute infinite-volume two-particles elastic scattering amplitudes. In the L\"uscher's finite-volume approach, the quantization condition is derived analytically and it is then used to establish a mathematical connection between the finite-volume two-particles energy levels (that are discrete and that can be extracted from suitably chosen lattice correlators) and the elastic scattering phase shifts.  In another fundamental paper~\cite{Lellouch:2000pv}, Lellouch and L\"uscher have then extended the finite-volume formalism to the calculation of matrix elements of two-particles states below the relevant inelastic threshold. Subsequently, the original derivation of L\"uscher's quantization condition has been generalized to the case of multiple channels of two-particle states~\cite{Rummukainen:1995vs, Bedaque:2004kc, Kim:2005gf, He:2005ey, Christ:2005gi, Lage:2009zv, Bernard:2010fp, Doring:2011vk, Doring:2011nd, Luu:2011ep, Doring:2011ip, Fu:2011xz, Doring:2012eu, Hansen:2012tf, Briceno:2012yi, Bernard:2012bi, Guo:2012hv, Gockeler:2012yj, Leskovec:2012gb, Agadjanov:2013kja, Briceno:2014oea, Grabowska:2021xkp, Lin:2001ek, Meyer:2011um, Briceno:2014uqa, Agadjanov:2014kha, Briceno:2015csa} (this allows to study QCD processes in which e.g.\ a two-pion state can rescatter into a two-kaon state) and more recently also to three-particles states~\cite{Detmold:2008fn, Polejaeva:2012ut, Briceno:2012rv, Hansen:2014eka, Hansen:2015zga, Briceno:2017tce, Hammer:2017uqm, Hammer:2017kms, Guo:2017crd, Guo:2017ism, Guo:2018xbv, Briceno:2018aml, Blanton:2019igq, Briceno:2019muc, Romero-Lopez:2019qrt, Blanton:2020gmf, Blanton:2020jnm, Blanton:2020gha, Hansen:2020zhy, Guo:2020kph, Hansen:2019nir, Jackura:2019bmu, Rusetsky:2019gyk, Muller:2020wjo, Hansen:2021ofl, Muller:2021uur, Blanton:2021mih, Blanton:2021llb, Blanton:2021eyf, Muller:2022oyw, Bubna:2023oxo, Pang:2023jri, Draper:2023boj, Draper:2023xvu, Hansen:2024ffk, Baeza-Ballesteros:2024mii,Draper:2024qeh}. The resulting formalism in the case of three-particle states is so involved that it is hard to believe that further generalizations, that would allow to study phenomenologically interesting processes such as e.g. $B\mapsto \pi\pi$ (where the threshold for producing more than 30 pions is open), will ever be obtained or could have practical applicability.

In this work we approach the theoretical problem of the extraction of $S$-matrix elements from Euclidean correlators from a continuum, infinite-volume perspective. It turns out that scattering amplitudes can be approximated arbitrarily well by means of linear combinations of Euclidean correlators at discrete time separations and suitably smeared with respect to the spatial coordinates. We work under the assumption that these quantities have been computed numerically on the lattice and then extrapolated to the continuum and infinite-volume limits by properly quantifying the systematic errors associated with these extrapolations. In fact, the approximation formulae presented here are ready to be used in numerical calculations of non-perturbative QCD scattering amplitudes. Whether these formulae can lead to the calculation of scattering amplitudes with a satisfactory precision remains to be seen. A concrete numerical strategy, built on the numerical methods developed in refs.~\cite{Hansen:2019idp,Bulava:2021fre} and then successfully applied in the non-perturbative calculation of inclusive hadronic quantities in refs.~\cite{Gambino:2022dvu, ExtendedTwistedMassCollaborationETMC:2022sta, Barone:2023tbl, Bonanno:2023ljc, Frezzotti:2023nun, Bonanno:2023thi, Evangelista:2023fmt, Alexandrou:2024gpl}, is sketched in section~\ref{sec:remarks} and will be discussed in details in future publications. We concentrate here on the theoretical issues associated with the derivation of our results.

When talking about scattering in QFT, it is useful to contrast the theories developed by Lehmann, Symanzik and Zimmermann~\cite{Lehmann:1954rq} (LSZ) on the one hand, and by Haag and Ruelle~\cite{Haag:1958vt,Ruelle1962} on the other hand. While the LSZ formalism is well known to particle physicists, and commonly adopted in practical calculations, the Haag-Ruelle formalism is much less known and rarely, if ever, used in calculations. The success of the LSZ formalism stems from the fact that scattering amplitudes are expressed in terms of time-ordered correlators in Minkowski space, which can be easily calculated in perturbation theory by means of Feynman diagrams. Nevertheless, Haag-Ruelle theory is more fundamental than the LSZ theory, in the sense that it allows to define asymptotic multi-particle states and not merely their matrix elements, essentially by giving a rigorous and non-perturbative meaning to the textbook expressions $\Omega_{\pm}=\lim_{t \to \pm\infty}e^{itH_0}e^{-itH}$ for the M{\o}ller operators (in terms of which, the scattering operator is given by $S=\Omega_+^\dagger \Omega_-$). In fact, the LSZ reduction formulae have to be derived starting from Haag-Ruelle scattering theory and this has rigorously been done in ref.~\cite{cmp/1103758732}.

Our approximation formulae are derived within the framework of Haag-Ruelle theory. This choice is dictated by a number of clear advantages of Haag-Ruelle over LSZ scattering theory. Firstly, scattering amplitudes are related to Euclidean correlators more directly via the Haag-Ruelle formalism: scattering amplitudes are natively expressed as spectral densities smeared with Schwartz kernels, while Euclidean correlators are Laplace transforms of the same spectral densities. In other words, Haag-Ruelle theory provides the shortest path to our approximation formula. Secondly, the full power of the Haag-Ruelle theory allows to determine the scaling of the systematic error induced by our approximation formula, as a function of a couple of approximation parameters. Lastly, Haag-Ruelle theory allows to derive a family of approximation formulae in terms of certain auxiliary functions which are largely arbitrary. This feature, that may look like a nuisance at first sight, is on the contrary a dial that can be used to minimize the systematic and statistical errors in realistic numerical calculations.

In terms of general goals, this paper presents strong similarities with previous work of Barata and Fredenhagen. In their forward-looking and inspiring paper~\cite{Barata:1990rn}, Barata and Fredenhagen have addressed some crucial aspects of the problem of extracting $S$-matrix elements from Euclidean correlators in a lattice-discretized theory. In particular, they have shown that asymptotic states and, hence, scattering amplitudes can be rigorously defined in the lattice-discretized theory and can be approximated arbitrarily well by linear combinations of Euclidean correlators. This approximation is, at least in principle, calculable by means of lattice simulations. In the original Barata-Fredenhagen construction, the approximant does not necessarily have a well-defined continuum limit: the error on the approximation must be made vanishingly small before one can attempt a continuum extrapolation. Our philosophy is complementary: we show here that, even in the continuum theory, the scattering amplitude can be approximated by means of linear combinations of spatially-smeared Euclidean correlators. In order to achieve these results we had to cope with some fairly challenging mathematical subtleties, not encountered by Barata and  Fredenhagen, originating from the fact that  continuum fields are operator-valued distributions while lattice fields are bounded operators (at least in gauge theories coupled to fermions). For this technical reason, while our approximation formulae can be considered a generalization of those previously obtained by Barata and Fredenhagen, the mathematical proofs of our theorems require rather different techniques and, therefore, represent an original and (in our opinion) important result on their own.

In ref.~\cite{Bulava:2019kbi}, Bulava and Hansen studied the problem of the extraction of scattering amplitudes from Euclidean correlators by starting from the LSZ reduction formula. The main result of their investigation, derived under certain physically-plausible mathematical assumptions, are distributional expressions for scattering amplitudes given in terms of spectral densities convoluted with energy propagators, i.e. with the distributions $1/(E\pm i0^+)$. Bulava and Hansen then envisage replacing the energy correlators with Cauchy's smearing kernels, i.e. with $1/(E\pm i\sigma)$ at finite values of the smearing parameter $\sigma$, in order to be able to apply the numerical techniques developed in ref.~\cite{Hansen:2019idp} for the extraction of smeared spectral densities from Euclidean correlators. In contrast, the smearing kernels appearing in our expressions for the scattering amplitudes, which naturally arise from Haag-Ruelle scattering theory, are much more general with respect to the ones envisaged by Bulava and Hansen and, by construction, are fully compatible with the axiomatic framework. Moreover, we anticipate that the original Haag-Ruelle construction can be generalized and, as a result, the class of smearing kernels can be significantly enlarged, thus providing more options which may be interesting in view of future numerical applications. These generalizations will be the subject of future work.

The paper is structured as follows. In section~\ref{sec:framework} we briefly review the Haag-Ruelle scattering theory by focusing on the main results, the axiomatic definition of the asymptotic states, on which our construction is then built. Section~\ref{sec:scattering} presents the main results of this paper, organized in various subsections. In subsection~\ref{subsec:tantalo} we derive a first useful representation of scattering amplitudes in terms of smeared Wightman functions. Scattering amplitudes are then rewritten in terms of spectral densities in subsection~\ref{subsec:spectral}. Spectral densities are related to Euclidean correlators in subsection~\ref{subsec:euclidean}.
Finally, the scattering amplitude is written in terms of Euclidean correlators at discrete time separations in subsection~\ref{subsec:approx}. In section~\ref{sec:matrix}, the construction is generalized to matrix elements of local operators between incoming or outgoing asymptotic states. In section~\ref{sec:remarks} we summarize and discuss our results. The technical appendices contain the detailed mathematical proofs of the results presented and discussed in the main text.

\section{Theoretical framework}\label{sec:framework}
% !TEX root = paper.tex
% !TeX spellcheck = en-US

We work in the framework of Wightman axioms~\cite{Wightman:1956zz} and Haag-Ruelle scattering theory~\cite{Haag:1958vt,Ruelle1962} (see also~\cite{Haag:1996hvx,Duncan:2012aja} for a textbook introduction to the subject). For simplicity, we consider a Quantum Field Theory which contains only one particle with mass $m>0$ and zero spin. Therefore the the squared mass operator $P^2$ has a unique non-negative discrete eigenvalue $m^2$, while the continuum part of the spectrum is given by $[4m^2,+\infty)$. While the existence of a mass gap is an essential assumption behind Haag-Ruelle scattering theory, the formalism can be easily extended to theories with more stable particles.

Haag-Ruelle scattering theory allows to construct creation and annihilation operators for asymptotic particles. The creation and annihilation operators for outgoing particles, $a_\sout(\vec{p})^\dag$ and $a_\sout(\vec{p})$ respectively, satisfy the standard relations:
\begin{gather}
    [ a_\sout(\vec{p}) , a_\sout(\vec{q})^\dag ] = (2\pi)^3 \delta^3( \vec{p} - \vec{q} )
    \ , \qquad
    [ a_\sout(\vec{p}) , a_\sout(\vec{q}) ] = 0
    \ , \\
    [ H , a_\sout(\vec{p})^\dag ] = E(\vec{p}) \, a_\sout(\vec{p})^\dag
    \ , \qquad
    [ \vec{P} , a_\sout(\vec{p})^\dag ] = \vec{p} \, a_\sout(\vec{p})^\dag
    \ , \\
    a_\sout(\vec{p}) \rvac = 0
    \ ,
\end{gather}
where $E(\vec{p})=\sqrt{m^2+\vec{p}^2}$ is one-particle energy, $H$ and $\vec{P}$ are the Hamiltonian and momentum operators and $\rvac$ denotes the vacuum state, i.e. the ground state of the Hamiltonian $H$. Similar relations are satisfied by the creation and annihilation operators for ingoing particles, $a_\sin(\vec{p})^\dag$ and $a_\sin(\vec{p})$, respectively. To be precise, $a_\sin(\vec{p})$ and $a_\sout(\vec{p})$ are operator-valued distributions. Given a Schwartz wave function $\check{f}(\vec{p})$, we define
\begin{gather}
    a_\sin(\check{f}) = \int \frac{d^3\vec{p}}{(2\pi)^3} \check{f}^*(\vec{p}) a_\sin(\vec{p})
    \ , \qquad
    a_\sout(\check{f}) = \int \frac{d^3\vec{p}}{(2\pi)^3} \check{f}^*(\vec{p}) a_\sout(\vec{p})
    \ .
\end{gather}
The antilinear dependence on $\check{f}$ is necessary in order to interpret $a_\sin(\check{f})^\dag \rvac$ and $a_\sout(\check{f})^\dag \rvac$ as one-particle states with wave function $\check{f}(\vec{p})$. Throughout this paper, we will consider only smooth wave functions $\check{f}(\vec{p})$ with compact support (see appendix~\ref{app:compact} for explicit examples).

In order to construct asymptotic states, one introduces operators with the general form
\begin{gather}
    A(f,t) = \int \frac{d^4 p}{(2\pi)^4} e^{-i [ p_0 - E(\vec{p}) ] t} f(p)^* \tilde{\phi}(p)
    \ ,
    \label{eq:intro:A}
\end{gather}
where $\tilde{\phi}(p)$ is a local field in momentum space, i.e. the Fourier transform of a local field $\phi(x)$ in the Heisenberg picture:
\begin{gather}
    \tilde{\phi}(p) = \int d^4x \, e^{ipx} \phi(x) \ ,
\end{gather}
whose normalization is chosen in such a way that
\begin{gather}
    \lvac a_\sin(\vec{p}) \phi(x)^\dag \rvac
    =
    \lvac a_\sout(\vec{p}) \phi(x)^\dag \rvac
    =
    e^{i E(\vec{p}) x_0 - i \vec{p} \vec{x}}
    \ .
\end{gather}
The central result of the Haag-Ruelle scattering theory is the existence of the scattering states and the fact that they can be constructed by means of the following strong limits:
\begin{subequations}
\label{eq:intro:exstates}
\begin{gather}
    \lim_{t \to -\infty} A(f_N,t)^\dag \cdots A(f_1,t)^\dag \rvac
    =
    a_\sin(\check{f}_N)^\dag \cdots a_\sin(\check{f}_1)^\dag \rvac
    \ , \\
    \lim_{t \to +\infty} A(f_N,t)^\dag \cdots A(f_1,t)^\dag \rvac
    =
    a_\sout(\check{f}_N)^\dag \cdots a_\sout(\check{f}_1)^\dag \rvac
    \ ,
\end{gather}
\end{subequations}
provided that the functions $f_A$ satisfy two conditions: \textit{(1)} the closed support of $f_A$ intersects the spectrum of energy-momentum operator $P$ only on the mass-shell $p^2=m^2$, \textit{(2)} the restriction of $f_A$ on the mass-shell gives the wave function, i.e. $f_A(E(\vec{p}),\vec{p}) = \check{f}_A(\vec{p})$. The limits in eq.~\eqref{eq:intro:exstates} are reached with an error in norm of order $|t|^{-1/2}$~\cite{Haag:1958vt} in the general case. In the special case of non-overlapping velocities, i.e. if the set of velocities
\begin{gather}
    V_A = \{ \nabla E(\vec{p}) \text{ s.t. } \vec{p} \in \supp \check{f}_A \}
\end{gather}
are pairwise disjoint, the limits in eq.~\eqref{eq:intro:exstates} are reached with an error in norm that vanishes faster than any inverse power of $|t|$~\cite{cmp/1103758732}. Notice that, throughout this paper, the symbol $\supp f$ denotes the \textit{closed support} of $f$.

In the case of wave functions $\check{f}_A(\vec{p})$ with compact support one can choose the functions $f_A$ of the following form
\begin{gather}
    f_A(p)
    =
    \zeta_A(p_0 {-} E(\vec{p})) \check{f}_A(\vec{p})
    \ ,
    \label{eq:intro:f}
\end{gather}
where $\zeta_A(\omega)$ are smooth functions with compact support satisfying $\zeta_A(0)=1$.\footnote{
Operators of the general form~\eqref{eq:intro:A} were not present in Haag's original work~\cite{Haag:1958vt} and were given e.g. by Hepp in~\cite{cmp/1103758732}. We notice that, even though not manifest, the operators with $f_A(p)$ given by~\eqref{eq:intro:f} are indeed a specially case of the operators introduced originally by Haag~\cite{Haag:1958vt}:
\begin{gather*}
    A(f,t) = i \int d^3 \vec{x} \, K(t,\vec{x}) \overleftrightarrow{\partial_0} q(t,\vec{x})
    \ , \qquad
    K(x) = \int \frac{d^3 \vec{p}}{(2\pi)^3} e^{i E(\vec{p}) x_0 - i \vec{p} \vec{x}} \check{f}^*(\vec{p})
    \ .
\end{gather*}
The so-called quasi-local field $q(x)$ is defined by means of its Fourier transform
\begin{gather*}
    \tilde{q}(p) = \frac{ \tilde{g}(p) \zeta^*(p_0 {-} E(\vec{p})) }{ p_0 + E(\vec{p}) } \tilde{\phi}(p)
    \ ,
\end{gather*}
where $\tilde{g}(p)$ is a smooth function with the following properties: \textit{(1)} its compact support intersects the spectrum of energy-momentum operator $P$ only on the mass-shell $p^2=m^2$, and \textit{(2)} $\tilde{g}(p)$ is equal to one in the support of the function $p \mapsto \zeta(p_0 {-} E(\vec{p})) \check{f}(\vec{p})$.
} The existence of the functions $\zeta_A$ is proved in appendix~\ref{app:haag} (see also appendix~\ref{app:compact} for an explicit example).

\section{Approximation of scattering amplitudes}\label{sec:scattering}

\subsection{A useful representation for scattering amplitudes}\label{subsec:tantalo}
Consider $M \ge 2$ incoming particles with wave functions $\check{f}_{A=1,\dots,N}(\vec{p})$ and $N \ge 2$ outgoing particles with wave functions $\check{f}_{A=M+1,\dots,M+N}(\vec{p})$. The scattering amplitude for the considered scattering event $\check{f}_1 , \dots , \check{f}_M \to \check{f}_{M+1} , \dots , \check{f}_{M+N}$ is given by
\begin{gather}
    S = \lvac
    \rprod_{A=M+1}^{M+N} a_\sout(\check{f}_A) \lprod_{A=1}^M a_\sin(\check{f}_A)^\dag \rvac
    \nonumber \\
    =
    \int \Bigg[ \prod_{A=1}^{M+N} \frac{d^3 \vec{p}_A}{(2\pi)^3} \check{f}_A^{(*)}(\vec{p}_A) \Bigg]
    \lvac \rprod_{A=M+1}^{M+N} a_\sout(\vec{p}_A) \lprod_{A=1}^M a_\sin(\vec{p}_A)^\dag \rvac
    \ ,
    \label{eq:tantalo:S}
\end{gather}
where the symbol $\check{f}_A^{(*)}$ stands for the wave function $\check{f}_A$ if $A \le M$ and its complex conjugate if $A > M$, the product symbol with a right (resp. left) arrow indicates that factors must be ordered by increasing index from left to right (resp. right to left). In order to have a non-vanishing scattering matrix, we require that some momenta $\bar{\vec{p}}_A$ exist such that $\check{f}_A(\bar{\vec{p}}_A) \neq 0$ and which satisfy the energy-momentum conservation conditions
\begin{gather}
    \sum_{A=1}^{M+N} \eta_A \bar{\vec{p}}_A = 0
    \ , \qquad
    \sum_{A=1}^{M+N} \eta_A E(\bar{\vec{p}}_A) = 0
    \ ,
\end{gather}
where $\eta_A=+1$ (resp. $\eta_A=-1$) if the index $A$ corresponds to an outgoing (resp. incoming) particle.

Haag-Ruelle theory yields the relevant asymptotic states as the $t \to \pm\infty$ limit (in strong sense) of the following states
\begin{subequations}
\label{eq:tantalo:Psi}
\begin{gather}
    | \Psi_\sin(t) \rangle = A(f_M,t)^\dag \cdots A(f_1,t)^\dag \rvac
    \ , \\
    | \Psi_\sout(t) \rangle = A(f_{M+N},t)^\dag \cdots A(f_{M+1},t)^\dag \rvac
    \ ,
\end{gather}
\end{subequations}
and the scattering amplitude as the $t \to +\infty$ limit of the transition probability
\begin{gather}
    \label{eq:tantalo:PsiPsi}
    \langle \Psi_\sout(t) | \Psi_\sin(-t) \rangle
    =
    \int \Bigg[ \prod_{A=1}^{M+N} \frac{d^4 p_A}{(2\pi)^4} f_A^{(*)}(p_A) \Bigg]
    e^{-i \sum_{A=1}^{M+N} [ p_{A,0} {-} E(\vec{p}_A) ] t}
    \tilde{W}(p)
    \ ,
\end{gather}
which involves the Wightman function in momentum space:
\begin{gather}
    \label{eq:tantalo:WS}
    \tilde{W}(p) =
    \lvac \tilde{\phi}(p_{M+1}) \cdots \tilde{\phi}(p_{M+N}) \tilde{\phi}(p_M)^\dag \cdots \tilde{\phi}(p_1)^\dag \rvac \ .
\end{gather}

Notice that, in eq.~\eqref{eq:tantalo:PsiPsi}, the complex exponential in the integrand oscillates more and more wildly as $t \to + \infty$, at least for large values of $\left| \sum_A [p_{A,0}{-}E(\vec{p}_A)] \right|$. In view of a possible approximation strategy, this feature in undesirable. The oscillatory behavior can be partially regulated with the following trick. We introduce two unit-integral Schwartz functions $h(s)$ and $\Phi(\tau)$, with the additional requirement that $\Phi(\tau)$ has closed support in $(0,+\infty)$. Given some $\sigma>0$ we define the integrated transition probability
\begin{gather}
    \mathcal{S}(\sigma) = \sigma \int dt \, ds \, \Phi(t\sigma) \, h(s) \, \Big\langle \Psi_\sout\left( \tfrac{t}{2}{-}s \right) \Big| \Psi_\sin\left( -\tfrac{t}{2}{-}s \right) \Big\rangle
    \ .
    \label{eq:tantalo:T-def}
\end{gather}

Translational invariance implies that the Wightman function is proportional to a delta of energy-momentum conservation~\cite{haag1992local,jost1965general}. Using this fact, which allows to set $\sum_{A=1}^{M+N} \eta_A p_{A,0}$ to zero in the integrand below, a straightforward calculation yields
\begin{gather}
    \mathcal{S}(\sigma)
    =
    \int \Bigg[ \prod_{A=1}^{M+N} \frac{d^4 p_A}{(2\pi)^4} f_A^{(*)}(p_A) \Bigg] \,
    \tilde{h}\left( \sum_{A=1}^{M+N} \eta_A E(\vec{p}_A) \right)
    \nonumber \\
    \phantom{\mathcal{S}(\sigma) = \int \Bigg[ \prod_{A=1}^{M+N} \frac{d^4 p_A}{(2\pi)^4} f_A^{(*)}(p_A) \Bigg]}
    \times \tilde{\Phi} \left( \frac{1}{2\sigma} \sum_{A=1}^{M+N} [ p_{A,0} {-} E(\vec{p}_A) ] \right) \,
    \tilde{W}(p)
    \ ,
    \label{eq:tantalo:T-int}
\end{gather}
which is written in terms of the Fourier transforms
\begin{gather}
    \tilde{h}(E) = \int ds \, e^{-i E s} h(s) \ ,
    \qquad
    \tilde{\Phi}(\omega) = \int d\tau \, e^{-i\omega \tau} \Phi(\tau) \ .
\end{gather}
We claim that
\begin{gather}
    \mathcal{S}(\sigma)
    \overset{\sigma \to 0^+}{=}
    S
    + O(\sigma^{r})
    \ ,
    \label{eq:tantalo:TS}
\end{gather}
where $r$ can be an arbitrary positive number in the case of non-overlapping velocities and $r=1/2$ in the general case. The proof of this statement is postponed to appendix~\ref{app:transition}.

Here we want to provide some insight into our construction. The role of the function $h$ can be understood by means of eq.~\eqref{eq:tantalo:T-int}. Notice that the condition that $h(s)$ has unit integral is equivalent to the condition $\tilde{h}(0)=1$. $\tilde{h}$ can be chosen to be arbitrarily peaked around zero, and even with compact support (see appendix~\ref{app:compact} for an explicit example), without affecting the $\sigma \to 0^+$ limit of $\mathcal{S}(\sigma)$. This means that regions of the integral in eq.~\eqref{eq:tantalo:T-int} characterized by arbitrarily small violations of the asymptotic-particle energy-conservation condition $\sum_{A=1}^{M+N} \eta_A E(\vec{p}_A)=0$ do not contribute to the $\sigma \to 0^+$ limit of $\mathcal{S}(\sigma)$, i.e. to the scattering amplitude, as it should be.

The role of the function $\Phi$ is better understood by means of eq.~\eqref{eq:tantalo:T-def} and it amounts to a pure mathematical trick. Let us we rewrite eq.~\eqref{eq:tantalo:T-def} as
\begin{gather}
    \mathcal{S}(\sigma) = \sigma \int dt \, \Phi(t\sigma) \, \mathcal{I}(t) = \int_0^\infty d\tau \, \Phi(\tau) \, \mathcal{I}\left(\frac{\tau}{\sigma}\right)
    \ ,
\end{gather}
where we have implicitly defined $\mathcal{I}(t)$ and we have used the fact that $\Phi(\tau)$ has closed support in $(0,+\infty)$. If we take now the $\sigma \to 0^+$ limit, we get
\begin{gather}
    \lim_{\sigma \to 0^+} \mathcal{S}(\sigma) = \int_0^\infty d\tau \, \Phi(\tau) \, \lim_{\sigma \to 0^+} \mathcal{I}\left(\frac{\tau}{\sigma}\right)
    =
    \mathcal{I}(+\infty) \int_0^\infty d\tau \, \Phi(\tau) = \mathcal{I}(+\infty)
    \ ,
\end{gather}
where we have used the fact that $\Phi(\tau)$ has unit integral. Therefore the integration against $\sigma \, \Phi(\sigma t)$ is just a mathematical trick which allows to trade the $t \to +\infty$ limit with the $\sigma \to 0^+$ limit, while partially regulating the oscillatory exponential of eq.~\eqref{eq:tantalo:PsiPsi}. The property that $\Phi(\tau)$ has closed support in $(0,+\infty)$ essentially selects the desired time-ordering of the asymptotic states and implies that $\tilde{\Phi}(\omega)$ must be complex (see appendix~\ref{app:compact} for an explicit example).

\subsection{Relation between transition amplitudes and spectral densities}\label{subsec:spectral}
Even though not necessary, it is convenient to consider the connected scattering amplitude, denoted in general with a subscript $c$, and defined by replacing the expectation value in eq.~\eqref{eq:tantalo:S} with its connected part. The connected transition amplitudes $\langle \Psi_\sout(t) | \Psi_\sin(-t) \rangle_c$ and $\mathcal{S}_c(\sigma)$ are obtained by replacing the Wightman function in eqs.~\eqref{eq:tantalo:PsiPsi} and \eqref{eq:tantalo:T-int} with its connected part.\footnote{In the mathematical physics literature, connected Wightman distributions are often referred to as \textit{truncated} Wightman distributions.} Using the standard algebra that relates expectation values to connected expectation values, it is easy to show that eq.~\eqref{eq:tantalo:TS} remains valid also for the connected parts, i.e.
\begin{gather}
    S_c
    \overset{\sigma \to 0^+}{=}
    \mathcal{S}_c(\sigma)
    + O(\sigma^r)
    \ .
    \label{eq:spectral:TcSc}
\end{gather}

Let us have a deeper look at the connected Wightman function. Using the representation
\begin{gather}
    \tilde{\phi}(p) = \int d^4x \, e^{iPx} \phi(0) e^{i(p-P)x} \ ,
\end{gather}
in eq.~\eqref{eq:tantalo:WS} and the translational invariance of the vacuum, one can perform the integrations over the coordinates explicitly, obtaining a chain of delta functions. The connected Wightman function can thus be rewritten as a spectral density times the delta of energy-momentum conservation:
\begin{gather}
    \label{eq:spectral:W-rho}
    \tilde{W}_c(p) = (2\pi)^4 \delta^4( q_{M+N}{-}q_M )
    \\ \nonumber
    \phantom{\tilde{W}_c(p) =} \times
    \lvac
    \Bigg[
    \rprod_{A=M+1}^{M+N-1} \phi(0) (2\pi)^4 \delta^4(P{-}q_A)
    \Bigg]
    \phi(0)
    \Bigg[
    \lprod_{A=1}^M (2\pi)^4 \delta^4(P{-}q_A) \phi(0)^\dag
    \Bigg] \rvac_c
    \ ,
\end{gather}
and we have used the identification:
\begin{gather}
    q_{A \le M} = \sum_{B=1}^A p_B \ , \qquad q_{A > M} = \sum_{B=M+1}^A p_B
    \ .
    \label{eq:spectral:q}
\end{gather}
Eq.~\eqref{eq:spectral:W-rho} shows that the connected Wightman function vanishes if any of the $q_A$ is outside of the spectrum of the energy-momentum operator or even if any of the $q_A$ vanishes (since the connected part removes the vacuum contributions). In other words, the support of the connected Wightman function satisfies
\begin{gather}
    \label{eq:spectral:suppWc}
    \supp \tilde{W}_c \subseteq \left\{ (p_1,\dots,p_{M+N}) \text{ s.t. } q_{0,B} \ge E( \vec{q}_B ) \text{ for every } B \text{ and } q_{M+N}=q_M \right\}
    \ .
\end{gather}

While the variables $q_{A=1,\dots,M+N}$ are particularly suited to understand the support of the connected Wightman function, the proposed approximation strategy for the transition amplitude $\mathcal{S}_c(\sigma)$ is discussed more naturally in the variables $(\omega,p_{M+N,0},\vec{p})$, defined by introducing the $(M{+}N{-}1)$ \textit{off-shellness variables}:
\begin{gather}
    \omega_A =
    \begin{cases}
        \sum_{B=1}^A [ p_{B,0} - E(\vec{p}_B) ]
        \quad & \text{for } 1 \le A \le M \\[4pt]
        \sum_{B=M+1}^A [ p_{B,0} - E(\vec{p}_B) ]
        \quad & \text{for } M+1 \le A \le M+N-1
    \end{cases}
    \ .
    \label{eq:spectral:omega}
\end{gather}
The spectral density $\rho_c(\omega,\vec{p})$ is defined implicitly by the equation
\begin{gather}
    \label{eq:spectral:rho}
    \tilde{W}_c(p) = (2\pi)^4 \delta^4\left( \sum_{A=1}^{M+N} \eta_A p_A \right) \, \rho_c(\omega,\vec{p})
    \ ,
\end{gather}
where the identification \eqref{eq:spectral:omega} is used. A more explicit representation for the spectral density will be given in the next section.

We plug the above representation of the connected Wightman function in the connected part of the transition amplitude given by eq.~\eqref{eq:tantalo:T-int}. Then, we use the expression for the functions $f_A(p)$ given in eq.~\eqref{eq:intro:f}, we remove the delta of energy conservation by integrating over $p_{M+N,0}$ and we substitute the integration over the remaining $p_{A,0}$ variables with the integration over $\omega_A$ with $A<M+N$. These manipulations yield the connected transition amplitude as
\begin{gather}
\label{eq:spectral:S-sigma}
    \mathcal{S}_c(\sigma)
    =
    \int \Bigg[ \prod_{A=1}^{M+N} \frac{d^3 \vec{p}_A}{(2\pi)^3} \check{f}_A^{(*)}(\vec{p}_A) \Bigg] \,
    (2\pi)^3 \delta^3\left(  \sum_{A=1}^{M+N} \eta_A \vec{p}_A \right) \, \tilde{h}(\Delta(\vec{p}))
    \\ \nonumber \hspace{100pt} \times
    \int \Bigg[ \prod_{A=1}^{M+N-1} \frac{d \omega_A}{2\pi} \Bigg] \, K_\sigma(\omega,\Delta(\vec{p})) \rho_c(\omega,\vec{p})
    \ ,
\end{gather}
where we have defined the \textit{asymptotic energy violation},
\begin{gather}
    \Delta(\vec{p}) = \Delta(\vec{p}_1,\dots,\vec{p}_{M+N}) = \sum_{A=1}^{M+N} \eta_A E(\vec{p}_A) \ ,
    \label{eq:energy-violation}
\end{gather}
and the \textit{Haag-Ruelle kernel},
\begin{gather}
    \label{eq:spectral:K}
    K_\sigma(\omega,\Delta)
    = \tilde{\Phi} \left( \frac{2\omega_M - \Delta}{2\sigma} \right) \zeta_1\left( \omega_1 \right)
    \Bigg[ \prod_{A=2}^{M-1} \zeta_A\left( \omega_A {-} \omega_{A-1} \right) \Bigg]
    \zeta_{M}\left( \omega_M {-} \omega_{M-1} \right)
    \\ \nonumber \phantom{K_\sigma(\omega,\Delta) =}
    \times
    \zeta_{M+1}^*\left( \omega_{M+1} \right)
    \Bigg[ \prod_{A=M+2}^{M+N-1} \zeta_A^*\left( \omega_A {-} \omega_{A-1} \right) \Bigg]
    \zeta_{M+N}^*\left( \omega_M {-} \omega_{M+N-1} {-} \Delta \right)
    \ .
\end{gather}
The off-shellness variables $\omega$ have been defined in such a way that the Haag-Ruelle kernel $K_\sigma(\omega,\Delta(\vec{p}))$ depends on the spatial momenta only via the asymptotic energy violation $\Delta(\vec{p})$. Notice that eq.~\eqref{eq:spectral:S-sigma} allows to interpret the transition amplitude as a smeared version of the spectral density, where the wave functions and the Haag-Ruelle kernel can be thought as smearing kernels.

\subsection{Relation between spectral densities and Euclidean \texorpdfstring{$n$-pt}{n-pt} functions}\label{subsec:euclidean}
We introduce the field in time-momentum representation at time $x_0=0$, i.e.
\begin{gather}
    \hat{\phi}(\vec{p}) = \int d^3\vec{x} \, e^{-i\vec{p}\vec{x}} \phi(0,\vec{x})
    \ .
\end{gather}
A few lines of algebra yield the following representation of the spectral density:
\begin{gather}
    \rho_c(\omega,\vec{p})=
    \lvac
    \Bigg[
    \rprod_{A=M+1}^{M+N-1}
    \hat{\phi}(\vec{p}_A)
    2\pi \delta\left( H - \omega_A - \!\!\! \sum_{B=M+1}^{A} \!\! E(\vec{p}_B) \right)
    \Bigg]
    \phi(0)
    \\ \nonumber \phantom{\rho_c(\omega,\vec{p})= \lvac}
    \times \Bigg[
    \lprod_{A=1}^{M}
    2\pi \delta\left( H - \omega_A - \sum_{B=1}^{A} E(\vec{p}_B) \right)
    \hat{\phi}(\vec{p}_A)^\dag
    \Bigg] \rvac_c
    \ .
\end{gather}
The Laplace transform of the spectral density with respect to the variables $\omega$ satisfies the relation
\begin{gather}
    (2\pi)^3 \delta^3 \left( \sum_{A=1}^{M+N} \eta_A \vec{p}_A \right)
    \int \Bigg[ \prod_{A=1}^{M+N-1} \frac{d\omega_A}{2\pi} \, e^{-s_A \omega_A} \Bigg] \, \rho_c(\omega,\vec{p})
    =
    \Upsilon(s,\vec{p}) \hat{C}_c(s,\vec{p})
    \ ,
    \label{eq:euclidean:laplace}
\end{gather}
where we have introduced the connected Euclidean $(N{+}M)$-point functions in time-momentum representation:
\begin{gather}
    \hat{C}_c(s;\vec{p})
    =
    \lvac
    \Bigg[
    \rprod_{A=M+1}^{M+N-1}
    \hat{\phi}(\vec{p}_A) e^{-s_A H}
    \Bigg]
    \hat{\phi}(\vec{p}_{M+N})
    \Bigg[
    \lprod_{A=1}^{M}
    e^{-s_A H}
    \hat{\phi}(\vec{p}_A)^\dag
    \Bigg] \rvac_c
    \ ,
    \label{eq:euclidean:C}
\end{gather}
and the auxiliary functions
\begin{gather}
    \Upsilon(s;\vec{p})
    =
    \exp\left\{
    \sum_{A=1}^M s_A \sum_{B=1}^{A} E(\vec{p}_B)
    + \sum_{A=M+1}^{M+N-1} s_A \sum_{B=M+1}^{A} E(\vec{p}_B)
    \right\}
    \ .
    \label{eq:euclidean:Upsilon}
\end{gather}

\subsection{General structure of the spectral density}\label{subsec:order}
Before attacking the problem of approximating the scattering amplitude, we need to discuss an important point of technical nature. On the basis of Wightman axioms alone, the Wightman functions in momentum space, and hence the spectral densities, can be arbitrarily singular tempered distributions. As we will see, even though the sought approximation exists for any spectral density which is compatible with Wightman axioms, in order to have a procedural way to construct such an approximation, one needs to know how singular the spectral density can be in the particular case of interest. One may be tempted to think that, in realistic theories such as QCD, the spectral density is a function rather than a distribution. This is clearly not the case even for two-point function spectral densities, in which one must expect in general contributions from delta functions corresponding to stable particles. In the case of spectral densities which are relevant for scattering processes, one must expect at least products of distributions of the type $1/(\omega_A \pm i0^\pm)$ which, in a perturbative setup, can be understood as advanced and retarded propagators of incoming and outgoing particles. This structure has been exploited by Bulava and Hansen~\cite{Bulava:2019kbi}. Once clarified that the spectral densities are expected to be true distributions in realistic cases, we turn to the problem of obtaining a representation  which highlights their singular nature. This can be done in different ways: we choose a strategy which is dictated primarily by the need to keep the presentation as simple as possible, but one should keep in mind that different strategies are certainly possible and may be worth exploring in the future.

We will say that a function $f(x)$ of a finite number of real variables is tempered if and only if a constant $r \ge 0$ exists such that $(1+\|x\|)^{-r} f(x)$ is integrable. Notice that a tempered function is, in particular, locally integrable. We will say that $f(x)$ is $L^2$-tempered if it is tempered and locally $L^2$. Any tempered distribution can be written as the sum of possibly higher-order distributional derivatives of $L^2$-tempered functions.\footnote{
This is a trivial consequence of theorem VI in chapter VII of~\cite{schwartz1950theorie}, together with the observation that $(1+\|x\|^2)^{k/2} f(x)$ is a tempered function for any continuous bounded function $f(x)$ and any $k \ge 0$. See also theorem V.10 in~\cite{reed2003methods}.
} Roughly speaking, more derivatives correspond to more singular distributions. We illustrate this fact here with a few simple examples:
\begin{subequations}
\begin{gather}
    \delta(x) = \frac{\partial}{\partial x} \theta(x)
    \ , \\
    \delta'(x) = \frac{\partial^2}{\partial x^2} \theta(x)
    \ , \\
    f(x) \delta(y) + \delta(x) g(y) = \frac{\partial}{\partial y} [f(x) \theta(y)] + \frac{\partial}{\partial x} [ \theta(x) g(y) ]
    \ , \\
    \delta^{(d)}(x) = \frac{\partial}{\partial x_1} \cdots \frac{\partial}{\partial x_d} [ \theta(x_1) \cdots \theta(x_d) ]
    \ , \\
    \frac{\text{PV}}{x} = \frac{\partial}{\partial x} \log |x|
    \ , \\
    \frac{1}{x \pm i 0^+} = \frac{\partial}{\partial x} [ \log |x| \mp i \pi \theta(x) ]
    \ , \\
    \frac{1}{(x \pm i 0^+)^2} = - \frac{\partial^2}{\partial x^2} [ \log |x| \mp i \pi \theta(x) ]
    \ .
\end{gather}
\end{subequations}

Let us apply this general property of tempered distribution to the spectral density. One can always find some $L^2$-tempered functions $R_{\alpha,\beta}(\omega,\vec{p})$ labeled by the non-negative integer indices $\alpha_A$ and $\beta_{A,k}$ with $A=1,\dots,M+N-1$ and $k=1,2,3$ such that, for every Schwartz function $\varphi(\omega,\vec{p})$, the following identity holds
\begin{gather}
    \label{eq:order:rho-R}
    \int \Bigg[ \prod_{A=1}^{M+N-1} \frac{d \omega_A d^3 \vec{p}_A}{(2\pi)^4} \Bigg] \varphi(\omega,\vec{p}) \rho_c(\omega,\vec{p})
    \\ \nonumber \qquad =
    \sum_{ \substack{\vphantom{\beta}\alpha \text{ s.t.} \\ \|\alpha\|_1 \le \mathfrak{N}_\omega} }
    \sum_{ \substack{\beta \text{ s.t.} \\ \|\beta\|_1 \le \mathfrak{N}_{\vec{p}}} }
    \int \Bigg[ \prod_{A=1}^{M+N-1} \frac{d \omega_A d^3 \vec{p}_A}{(2\pi)^4} \Bigg] \, R_{\alpha,\beta}(\omega,\vec{p}) D_{\omega}^\alpha D_\vec{p}^\beta \varphi(q)
    \ .
\end{gather}
In the above expression we use the standard multi-index notation:
\begin{gather}
    D_\omega^\alpha = \prod_{A=1}^{M+N-1} \left( \frac{\partial}{\partial \omega_{A}} \right)^{\alpha_{A}}
    \ , \qquad
    D_{\vec{p}}^\beta = \prod_{A=1}^{M+N-1} \prod_{k=1}^3 \left( \frac{\partial}{\partial p_{A,k}} \right)^{\beta_{A,k}}
    \ .
\end{gather}
Notice that, while the integral sign in the left-hand side of eq.~\eqref{eq:order:rho-R} is just a formal symbol denoting the application of a distribution to a test function, the integral in the right-hand side is a true Lebesgue integral. The representation~\eqref{eq:order:rho-R} is generally not unique. The integers $(\mathfrak{N}_\omega,\mathfrak{N}_{\vec{p}})$ will appear explicitly in the construction of the approximation of the scattering amplitude.

In the final part of this section, we sketch a speculative argument that can be used to find reasonable estimates for $\mathfrak{N}_\omega$ and $\mathfrak{N}_{\vec{p}}$. We consider the smeared connected Wightman function in coordinate space
\begin{gather}
    W^g_c(x) = \lvac \Bigg[ \rprod_{A=M+1}^{M+N} \phi_{g_A}(x_A) \Bigg] \, \Bigg[ \lprod_{A=1}^{M} \phi_{g_A}(x_A)^\dag \Bigg] \rvac_c
    \label{eq:spec:Wgc(x)}
    \ ,
\end{gather}
written in terms of the semared fields
\begin{gather}
   \phi_{g}(x) = \int d^4y \, g(x-y) \phi(y) \ ,
\end{gather}
where $g_A(x)$ are arbitrary Schwartz functions. While the Wightman function in coordinate space is a tempered distribution, the smeared Wightman function in coordinate space is a polynomially-bounded smooth function. It is natural to expect that the singularity structure of the connected Wightman function in momentum space is related to the long distance behavior of the smeared connected Wightman function in coordinate space. If this is the case, one should be able to bound $(\mathfrak{N}_\omega,\mathfrak{N}_{\vec{p}})$ by bounding the long distance behavior of $W^g_c(x)$. Such bounds can be presumably obtained by assuming an effective-field-theory description of the long-distance physics. We present a concrete argument of conjectural nature as illustration of this fact.

In the case of QCD, one can view Lattice QCD as an effective theory description of the long-distance physics of QCD. Theorem 1 in Barata and Fredenhagen~\cite{Barata:1990rn} implies that the following bound holds for the lattice-discretized smeared connected Wightman functions:
\begin{gather}
    \left| W^g_c(x) \right| \le C_q(g) \frac{ [ 1 + m \, d_t(x_0) ]^q }{ [ 1 + m \, d_s(\vec{x}) ]^{q-1} }\ ,
    \label{eq:spec:BFbound}
\end{gather}
where $q$ is an arbitrary non-negative number, $C_q(g)$ depends $q$ and the smearing functions $g$ but not on $x$, $d_s(\vec{x})$ is the diameter of the set of spatial points, i.e.
\begin{gather}
    d_s(\vec{x}) = \max_{A,B=1,\dots,M} \| \vec{x}_A - \vec{x}_B \|_2 \ ,
\end{gather}
and $d_t(x_0)$ is the maximum time separation between consecutive smeared operators as they appear in the right-hand side of eq.~\eqref{eq:spec:Wgc(x)}, i.e.
\begin{gather}
    d_t(x_0) = \max_{A=1,\dots,M+N-1} | \tau_A |
    \ ,
\end{gather}
with the definitions:
\begin{gather}
    \tau_{A<M} = x_{A,0}-x_{A+1,0}
    \ , \quad
    \tau_{A>M} = x_{A+1,0}-x_{A,0}
    \ , \quad
    \tau_M = x_{M,0}-x_{M+N,0}
    \ .
\end{gather}
Eq.~\eqref{eq:spec:BFbound} is essentially based on regularity properties which hold for the lattice-discretized theory, but not for a generic Wightman quantum field theory. If one assumes ultraviolet/infrared decoupling, i.e. the fact that the long-distance behavior of continuum QCD and Lattice QCD are essentially the same provided that the lattice spacing is small enough, it is natural to conjecture that a bound of the form given by eq.~\eqref{eq:spec:BFbound} should be valid in QCD as well. In appendix~\ref{app:order} we prove that, if the bound~\eqref{eq:spec:BFbound} holds for any $q \ge 0$ and some tempered distributions $C_q(g)$, then a representation of the type~\eqref{eq:order:rho-R} exists with
\begin{gather}
    \mathfrak{N}_\omega = 2(M+N) \ , \quad \mathfrak{N}_{\vec{p}} = 0 \ .
\end{gather}
We stress that, in a theory such as QCD, the validity of the assumptions behind this statement should be thoroughly scrutinized. In particular, we point out that the temperedness of the distribution $C_q(g)$ can not be motivated on the basis of the work of Barata and Fredenhagen and it is an \textit{ad hoc} assumption at this stage.

\subsection{Approximation in terms of Euclidean \texorpdfstring{$n$-pt}{n-pt} functions}\label{subsec:approx}
The goal of this section is to construct an approximation $\mathcal{S}_c(\sigma,\epsilon)$ of the connected scattering amplitude $S_c$ which is calculable in terms of a finite sampling in Euclidean time of certain Euclidean correlators. The quality of the approximation is controlled by two parameters, $\sigma$ and $\epsilon$. The connected scattering amplitude can then be recovered by taking the following double limit,
\begin{gather}
    S_c = \lim_{\sigma \to 0^+} \lim_{\epsilon \to 0^+} \mathcal{S}_c(\sigma,\epsilon) \ .
\end{gather}
We notice that the parameter $\sigma$ is the same one appearing in the construction of section~\ref{subsec:tantalo} and, morally, represents the inverse Minkowski time at which approximated asymptotic states are created. The desired approximation stems from a suitable approximation of the Haag-Ruelle kernel, defined in eq.~\eqref{eq:spectral:K}, whose accuracy is controlled by the parameter $\epsilon$ and which will be discussed in detail at the end of this section. The order of the two limits in the above formula can not be exchanged. However it is possible to take them simultaneously by e.g. fixing the ratio of the two parameters.

Let us see how to construct the approximant $\mathcal{S}_c(\sigma,\epsilon)$ explicitly. We choose some $\tau>0$ (playing the role of an elementary Euclidean time step), which we fix once and for all. We choose the function $\tilde{h}(\omega)$ introduced in subsection~\ref{subsec:tantalo} such that its support is inside $[-\bar{\Delta},\bar{\Delta}]$. Notice that the restriction on $\tilde{h}(\omega)$ is not essential, since $\tilde{h}(\omega)$ can be freely chosen as long as it is Schwartz and $\tilde{h}(0)=1$, and the scattering amplitude does not depend on this choice. In the following, $\mathfrak{N}_\omega$ and $\mathfrak{N}_{\vec{p}}$ are non-negative integers for which a representation of the spectral function of the form~\eqref{eq:order:rho-R} exists. We construct a polynomial $P_{\sigma,\epsilon}(z_1,\dots,z_{M+N-1},\Delta)$ which does not contain monomials of degree zero in any of the variables $z_A$ and satisfies the bound
\begin{gather}
    \sum_{ \substack{ \| \alpha \|_1 = \mathfrak{N}_\omega \\ 0 \le b \le \mathfrak{N}_{\vec{p}} } } \!\! \bar{\Delta}^{b}
    \int_{\mathbb{K}} \Bigg[ \prod_{A=1}^{M+N-1} \frac{d\omega_A}{2\pi} \Bigg] d\Delta \,
    e^{\tau \sum_A \omega_A}
    \left| D_\omega^\alpha \partial_\Delta^b
    \left[ K_\sigma(\omega,\Delta) - P_{\sigma,\epsilon}(e^{-\tau \omega},\Delta) \right] \right|^2 < \epsilon^2
    \ ,
    \label{eq:approx:P}
\end{gather}
where the approximation domain $\mathbb{K}$ is given by
\begin{gather}
    \mathbb{K} = [\bar{\omega}_1{-}\bar{\Delta},+\infty) \times \cdots \times [\bar{\omega}_{M+N-1}{-}\bar{\Delta},+\infty) \times [ -\bar{\Delta},\bar{\Delta} ]
    \label{eq:approx:Kbb}
    \ , \\
    \bar{\omega}_A = \inf \bigg\{ \omega_A \text{ s.t. } q_{0,B} \ge E(\vec{q}_B) , \, \vec{p}_B \in \supp \check{f}_B \text{ for every } B
    \nonumber \\ \hspace{3cm}
    \text{and } {\textstyle \sum_{B=1}^{M+N} \eta_B \vec{p}_B } = 0
    \text{ and } {\textstyle \left| \sum_{B=1}^{M+N} \eta_B E(\vec{p}_B) \right| } \le \bar{\Delta}
    \bigg\}
    \label{eq:approx:omegabar}
    \ .
\end{gather}
In the definition of $\bar{\omega}_A$, the identifications~\eqref{eq:spectral:q} and \eqref{eq:spectral:omega} are understood. A few comments are in order.
\begin{enumerate}

    \item We claim that $\bar{\omega}_A$ is finite. We prove this statements in appendix~\ref{app:approx:domain}. The quantity $\bar{\omega}_A$ and, hence, the domain $\mathbb{K}$ depend only on the input data of the problem, i.e. the wave functions, the parameter $\sigma$ and the arbitrary parameter $\bar{\Delta}$. In practice they can be calculated by solving a multi-dimensional minimization problem.

    \item We claim that, for any $\sigma>0$ and $\epsilon > 0$, a polynomial $P_{\sigma,\epsilon}(z,\Delta)$ which does not contain monomials of degree zero in any of the variables $z_A$ and satisfies eq.~\eqref{eq:approx:P} exists. We prove this statement in appendix~\ref{app:approx}. We will write:
    \begin{gather}
        P_{\sigma,\epsilon}(z,\Delta) = \sum_{n_1,\dots,n_{M+N-1} \ge 1} \sum_{b \ge 0} w^{\sigma,\epsilon}_{n,b} z_1^{n_1} \cdots z_{M+N-1}^{n_{M+N-1}} \Delta^b
        \ ,
        \label{eq:approx:w}
    \end{gather}
    where only a finite number of coefficients $w^{\sigma,\epsilon}_{n,b}$ are different from zero.

    \item The construction of the polynomial $P_{\sigma,\epsilon}(z,\Delta)$ is computationally straightforward. Notice that the left-hand side of eq.~\eqref{eq:approx:P} is a quadratic function of the real and imaginary parts of the coefficients $w^{\sigma,\epsilon}_{n,b}$ of the polynomial. Therefore, for a fixed degree, one can find the polynomial which minimizes the left-hand side of eq.~\eqref{eq:approx:P} simply by solving a linear system of equations, whose coefficients depend only on the input data of the problem. Then one can progressively increase the degree of the polynomial until eq.~\eqref{eq:approx:P} is satisfied. It is worth pointing out that eq.~\eqref{eq:approx:P} can be combined in a straightforward fashion with regularization procedures like the HLT method~\cite{Hansen:2019idp}.

    \item A certain arbitrariness exists in the choice of the left-hand side of eq.~\eqref{eq:approx:P}, for instance one could modify the weight in the integral to some extent. This possibility will not be explored here, but it may be relevant in practical applications.

\end{enumerate}

The sought approximation $\mathcal{S}_c(\sigma,\epsilon)$ of the scattering amplitude is obtained by replacing $K_\sigma(\omega,\Delta(\vec{p}))$ with its approximation $P_{\sigma,\epsilon}(e^{-\tau \omega},\Delta(\vec{p}))$ in the expression for the transition amplitude given by eq.~\eqref{eq:spectral:S-sigma}. Using the representation~\eqref{eq:approx:w} of the polynomial $P_{\sigma,\epsilon}(z,\Delta)$ one easily recognizes that the integrals over $\omega$ yield the Laplace transform of the spectral density. Using eq.~\eqref{eq:euclidean:laplace}, a straightforward calculation yields
\begin{gather}
    \mathcal{S}_c(\sigma,\epsilon) =
    \sum_{n_1,\cdots,n_{N+M-1} \ge 1} \sum_{b \ge 0} w^{\sigma,\epsilon}_{n,b}
    \nonumber \\
    \phantom{\sum_{n_1,\cdots,n_{N+M-1} \ge 1}}  \times
    \int \Bigg[ \prod_{A=1}^{M+N} \frac{d^3 \vec{p}_A}{(2\pi)^2} \check{f}_A^{(*)}(\vec{p}_A) \Bigg] \,
    \tilde{h}( \Delta(\vec{p}) )
     \Upsilon(\tau n;\vec{p}) \, [\Delta(\vec{p})]^b \, \hat{C}_c(\tau n;\vec{p})
    \ ,
    \label{eq:approx:S-sigma-eps}
\end{gather}
where $\hat{C}_c(\tau n;\vec{p})$ is the Euclidean correlator defined in eq.~\eqref{eq:euclidean:C} and $\Upsilon(\tau n;\vec{p})$ is a known function whose explicit expression is given in eq.~\eqref{eq:euclidean:Upsilon}, both calculated at Euclidean time separations $s_A = n_A \tau$. We claim that the error of the approximation $\mathcal{S}_c(\sigma,\epsilon)$ is bounded by:
\begin{gather}
    \left| \mathcal{S}_c(\sigma,\epsilon) - S_c \right| < a \epsilon + b_r \sigma^r \ ,
    \label{eq:approx:bound}
\end{gather}
where $r$ can be an arbitrary positive number in the case of non-overlapping velocities and $r=1/2$ in the general case, and $a$ and $b_r$ depend on all data of the problem but $\epsilon$ and $\sigma$. The proof of this statement is presented in appendix~\ref{app:approx:error}.

\section{Matrix elements of local fields}\label{sec:matrix}
The approximation strategy used for scattering amplitudes can be easily applied also to matrix elements of a generic local field $J(x)$ between asymptotic states:
\begin{gather}
    F_c = \lvac a_\sout(\check{f}_{M+1}) \cdots a_\sout(\check{f}_{M+N}) J(0) a_\sin(\check{f}_{M})^\dag \cdots a_\sin(\check{f}_{1})^\dag \rvac_c
    \ ,
    \label{eq:matrix:F}
\end{gather}
where either $M$ or $N$ can be chosen to be zero. This matrix element can be obtained as the infinite-time limit of matrix elements between Haag states, i.e.
\begin{gather}
    \langle \Psi_\sout(t) | J(0) | \Psi_\sin(-t) \rangle_c \overset{t \to +\infty}{=} F_c + O(t^{-r})
    \ ,
    \label{eq:matrix:PsiJPsi}
\end{gather}
where we use the definitions given in eq.~\eqref{eq:tantalo:Psi}. We comment on the subtleties of eq.~\eqref{eq:matrix:PsiJPsi} in the context of axiomatic quantum field theory in appendix~\ref{app:matrix}, and we use it as our starting point.

We introduce a Schwartz function $\Phi(\tau)$ with unit integral and closed support contained in $(0,+\infty)$, and define:
\begin{gather}
    \mathcal{F}_c(\sigma) = \sigma^2 \int dt_1 \, dt_2 \, \Phi(t_1\sigma) \, \Phi(-t_2\sigma) \, \langle \Psi_\sout(t_1) | J(0) | \Psi_\sin(t_2) \rangle_c
    \ .
    \label{eq:matrix:Fsigma}
\end{gather}
A straightforward generalization of the arguments presented in section~\ref{subsec:tantalo} and appendix~\ref{app:transition} allows to prove
\begin{gather}
    \mathcal{F}_c(\sigma)
    \overset{\sigma \to 0^+}{=}
    F_c
    + O(\sigma^{r})
    \ ,
    \label{eq:matrix:Fsigma-limit}
\end{gather}
where $r$ can be an arbitrary positive number in the case of non-overlapping velocities and $r=1/2$ in the general case.
An explicit expression for $\mathcal{F}_c(\sigma)$ can be worked out,
\begin{gather}
    \mathcal{F}_c(\sigma)
    =
    \int \Bigg[ \prod_{A=1}^{M+N} \frac{d \omega_A d^3 \vec{p}_A}{(2\pi)^4} \check{f}_A^{(*)}(\vec{p}_A) \Bigg] \, K_\sigma(\omega) \rho_c^J(\omega,\vec{p})
    \ .
    \label{eq:matrix:Fsigma-int}
\end{gather}
The spectral density now depends on $(M{+}N)$ off-shellness variables $\omega$ and $(M{+}N)$ spatial momenta $\vec{p}$, and contains the insertion of $J(0)$:
\begin{gather}
    \label{eq:matrix:rho}
    \rho_c^J(\omega,\vec{p})=
    \lvac
    \Bigg[
    \rprod_{A=M+1}^{M+N}
    \hat{\phi}(\vec{p}_A)
    2\pi \delta\left( H - \omega_A - \!\!\! \sum_{B=M+1}^{A} E(\vec{p}_B) \right)
    \Bigg]
    J(0)
    \\ \phantom{\rho_c(\omega,\vec{p})= \lvac}
    \times \Bigg[
    \lprod_{A=1}^{M}
    2\pi \delta\left( H - \omega_A - \sum_{B=1}^{A} E(\vec{p}_B) \right)
    \hat{\phi}(\vec{p}_A)^\dag
    \Bigg] \rvac_c
    \ .
\end{gather}
The Haag-Ruelle kernel is given by
\begin{gather}
    \label{eq:matrix:K}
    K_\sigma(\omega)
    =
    \tilde{\Phi} \left( \frac{\omega_M}{\sigma} \right) \zeta_1\left( \omega_1 \right)
    \Bigg[ \prod_{A=2}^{M} \zeta_A\left( \omega_A {-} \omega_{A-1} \right) \Bigg]
    \\ \phantom{K_\sigma(\omega) =} \nonumber
    \times
    \tilde{\Phi} \left( \frac{\omega_{M+N}}{\sigma} \right)
    \zeta_{M+1}^*\left( \omega_{M+1} \right)
    \Bigg[ \prod_{A=M+2}^{M+N} \zeta_A^*\left( \omega_A {-} \omega_{A-1} \right) \Bigg]
    \ .
\end{gather}

The matrix element $F_c$ can be approximated by the quantity
\begin{gather}
    \label{eq:matrix:Fsigmaeps}
    \mathcal{F}_c(\sigma,\epsilon) =
    \sum_{n_1,\cdots,n_{N+M} \ge 1} w^{\sigma,\epsilon}_{n}
    \int \Bigg[ \prod_{A=1}^{M+N} \frac{d^3 \vec{p}_A}{(2\pi)^2} \check{f}_A^{(*)}(\vec{p}_A) \Bigg] \,
    \Upsilon(\tau n;\vec{p}) \, \hat{C}_c(\tau n;\vec{p})
    \ ,
\end{gather}
which is written in terms of the Euclidean correlators in time-momentum representation with an insertion of $J(0)$,
\begin{gather}
    \hat{C}_c^J(s;\vec{p})
    =
    \lvac
    \Bigg[
    \rprod_{A=M+1}^{M+N}
    \hat{\phi}(\vec{p}_A) e^{-s_A H}
    \Bigg]
    J(0)
    \Bigg[
    \lprod_{A=1}^{M}
    e^{-s_A H}
    \hat{\phi}(\vec{p}_A)^\dag
    \Bigg] \rvac_c
    \ ,
    \label{eq:matrix:C}
\end{gather}
and of the analytically-known functions
\begin{gather}
    \Upsilon(s;\vec{p})
    =
    \exp\left\{
    \sum_{A=1}^M s_A \sum_{B=1}^{A} E(\vec{p}_B)
    + \sum_{A=M+1}^{M+N} s_A \sum_{B=M+1}^{A} E(\vec{p}_B)
    \right\}
    \ .
    \label{eq:matrix:Upsilon}
\end{gather}
The coefficients $w^{\sigma,\epsilon}_{n}$ are the coefficients of a polynomial $P_{\sigma,\epsilon}(z)$
\begin{gather}
    P_{\sigma,\epsilon}(z) = \sum_{n_1,\dots,n_{M+N} \ge 1}  w^{\sigma,\epsilon}_{n} z_1^{n_1} \cdots z_{M+N}^{n_{M+N}}
    \ ,
    \label{eq:matrix:w}
\end{gather}
which satisfies the following approximation condition:
\begin{gather}
    \sum_{ \| \alpha \|_1 = \mathfrak{N}_\omega }
    \int_{\mathbb{K}} \Bigg[ \prod_{A=1}^{M+N} \frac{d\omega_A}{2\pi} \Bigg] \,
    e^{\tau \sum_A \omega_A}
    \left| D_\omega^\alpha
    \left[ K_\sigma(\omega) - P_{\sigma,\epsilon}(e^{-\tau \omega}) \right] \right|^2 < \epsilon^2
    \ ,
    \label{eq:matrix:approx-condition}
\end{gather}
where the approximation domain $\mathbb{K}$ is given by
\begin{gather}
    \mathbb{K} = [\bar{\omega}_1{-}\bar{\Delta},+\infty) \times \cdots \times [\bar{\omega}_{M+N}{-}\bar{\Delta},+\infty)
    \label{eq:matrix:Kbb}
    \ , \\
    \bar{\omega}_A = \inf \bigg\{ \omega_A \text{ s.t. } q_{0,B} \ge E(\vec{q}_B) , \, \vec{p}_B \in \supp \check{f}_B \text{ for every } B
    \bigg\}
    \label{eq:matrix:omegabar}
    \ ,
\end{gather}
and $\bar{\Delta}$ is an arbitrary positive number. In the definition of $\bar{\omega}_A$, the identifications~\eqref{eq:spectral:q} and
\begin{gather}
    \omega_A =
    \begin{cases}
        \sum_{B=1}^A [ p_{B,0} - E(\vec{p}_B) ]
        \quad & \text{for } 1 \le A \le M \\[4pt]
        \sum_{B=M+1}^A [ p_{B,0} - E(\vec{p}_B) ]
        \quad & \text{for } M+1 \le A \le M+N
    \end{cases}
    \label{eq:matrix:omega}
\end{gather}
are understood. In writing eq.~\eqref{eq:matrix:approx-condition}, we assume that a representation for the spectral density of the type given in eq.~\eqref{eq:order:rho-R} exists, with $\rho_c(\omega,\vec{p})$ replaced by $\rho_c^J(\omega,\vec{p})$. The parameter $\mathfrak{N}_\omega$ in eq.~\eqref{eq:order:rho-R} also appears in eq.~\eqref{eq:matrix:approx-condition}. We claim that the error of the approximation $\mathcal{F}_c(\sigma,\epsilon)$ is bounded by:
\begin{gather}
    \left| \mathcal{F}_c(\sigma,\epsilon) - F_c \right| < a \epsilon + b_r \sigma^r \ ,
\end{gather}
for some constants $a$ and $b_r$ which depend on all data of the problem but $\epsilon$ and $\sigma$. The proof of this statement is completely analogous to the corresponding one for the scattering amplitude.

\section{Summary and remarks}\label{sec:remarks}
In this paper, which elaborates on previous ideas and techniques developed by Barata and Fredenhagen in~\cite{Barata:1990rn}, we prove that scattering amplitudes can be approximated by means of eq.~\eqref{eq:approx:S-sigma-eps} as a linear combination of Euclidean correlators at discrete time separations, suitably smeared with respect to the spatial coordinates. A similar result is obtained also for matrix elements of local operators with respect to asymptotic states, see eq.~\eqref{eq:matrix:Fsigmaeps}.

Our approximation formulae constitute an interesting mathematical result per se, which we obtain in the context of Haag-Ruelle scattering theory. More importantly, we provide an algorithmic procedure to construct these approximations, which is surprisingly simple in spite of the mathematical and technical subtleties needed to derive it, and which can be used as the blueprint for a concrete numerical procedure, e.g. in the context of lattice QCD simulations. In fact, our approach turns the numerically ill-posed problem of analytically continuing Euclidean correlators to real time into a merely ill-conditioned problem of constructing sophisticated linear combinations. In view of future applications, we want to present some general comments on various aspects of our method, as well as indicate possible directions for improvements and extensions.

We comment first on the structure of the approximant $\mathcal{S}_c(\sigma,\epsilon)$ for the scattering amplitude $S_c$, given in eq.~\eqref{eq:approx:S-sigma-eps} and which we rewrite for the reader's convenience:
\begin{gather}
    \mathcal{S}_c(\sigma,\epsilon) =
    \sum_{n_1,\cdots,n_{N+M-1} \ge 1} \sum_{b \ge 0}\ w^{\sigma,\epsilon}_{n,b}\ \mathcal{C}(n,b) \ ,
    \nonumber \\
    \mathcal{C}(n,b) =
    \int \Bigg[ \prod_{A=1}^{M+N} \frac{d^3 \vec{p}_A}{(2\pi)^3} \check{f}_A^{(*)}(\vec{p}_A) \Bigg] \,
    \tilde{h}( \Delta(\vec{p}) )
     \Upsilon(\tau n;\vec{p}) \, [\Delta(\vec{p})]^b \, \hat{C}_c(\tau n;\vec{p})
    \ .
    \label{eq:remark}
\end{gather}
Before applying this formula, one needs to calculate the mass of the stable particles involved in the scattering process of interest, as well as the energy gap to the closest stable particle or multi-particle state with the same quantum numbers as the scattering particles. For instance, if one wants to calculate proton scattering amplitudes, one would need to know the proton mass as well as the pion mass, since this determines the lowest multi-particle threshold in the channel determined by baryon number equal to one. We assume that this step has been executed with standard lattice QCD techniques. Besides masses and energy gaps, the dynamics of the process is entirely encoded in the Euclidean correlator $\hat{C}_c(\tau n;\vec{p})$. All the other elements of the approximation formula are fixed by the kinematics of the process which is known a priori. The functions $\Upsilon(\tau n;\vec{p})$ and $\Delta(\vec{p})$ are explicitly given in eq.~\eqref{eq:euclidean:Upsilon} and eq.~\eqref{eq:energy-violation} respectively. The function $\tilde{h}( \Delta(\vec{p}) )$ can be conveniently chosen to improve the quality of the approximation of the scattering amplitude, as discussed in the paragraphs around eq.~\eqref{eq:tantalo:T-def}. The numerical coefficients $w^{\sigma,\epsilon}_{n,b}$ are obtained, for any fixed degree of the polynomial defined in eq.~\eqref{eq:approx:w}, by solving the linear set of equations corresponding to the minimization of the left-hand side of eq.~\eqref{eq:approx:P}. The order of the derivative appearing in eq.~\eqref{eq:approx:P} is related to the singularity structure of the spectral density and, therefore, has to be determined for each process. In subsection~\ref{subsec:order} we propose a argument of conjectural nature to estimate the order of the derivatives needed in the approximation formula. The construction of the approximant $\mathcal{S}_c(\sigma,\epsilon)$ depend on a number of auxiliary functions which are largely arbitrary. Concrete examples for such functions are provided in  appendix~\ref{app:compact}.

Lattice simulations introduce various sources of error, most notably discretization effects, finite-volume effects and statistical noise. When a finite volume $L^3$ (e.g. with periodic boundary conditions in space) is considered, an estimator for the function $\mathcal{S}_c(\sigma,\epsilon)$, which approximates the scattering amplitude, is obtained by simply replacing the integral over the spatial momenta in the definition of the spatially-smeared correlator $\mathcal{C}(n,b)$ with the corresponding sum over discrete momenta, i.e.
\begin{gather}
    \mathcal{C}(n,b) =
    \Bigg[ \prod_{A=1}^{M+N} \frac{1}{L^3} \sum_{\vec{p}_A} \check{f}_A^{(*)}(\vec{p}_A) \Bigg] \,
    \tilde{h}( \Delta(\vec{p}) )
     \Upsilon(\tau n;\vec{p}) \, [\Delta(\vec{p})]^b \, \hat{C}_c(\tau n;\vec{p})
    \ .
\end{gather}
Only a finite number (albeit increasing with $L$) of discrete momenta contribute to these sums because the spatial wave functions have compact support. We want to stress that, since only a finite sampling of the correlator in the time variable is needed and the spatial wave functions are Schwartz, the infinite-volume limit of the estimator for $\mathcal{S}_c(\sigma,\epsilon)$ is approached exponentially fast, provided that the coefficients $w^{\sigma,\epsilon}_{n,b}$, which control the quality of the approximation, are kept constant as the volume is varied. Of course, this does not automatically mean that finite-volume corrections are small, and they should be expected to become larger and larger as $\epsilon$ and $\sigma$ are decreased.

When a non-zero lattice spacing is considered, $\mathcal{S}_c(\sigma,\epsilon)$ is discretized in a straightforward way if the lattice spacing divides $\tau$, while in the general case one can design some simple interpolation motivated by the fact that the Euclidean correlators in the continuum are analytic functions of the coordinates. One could also generalize our approximation strategy and introduce some smearing in time. In all cases, we expect that the continuum limit of the estimator can be understood in terms of the Symanzik effective theory or simple extensions of it. We stress that, in this approach, $\tau$ must be kept fixed in physical units. This is the main difference with respect to the Barata-Fredenhagen approach~\cite{Barata:1990rn} in which a sampling of the Euclidean correlator with $\tau$ equal to the lattice spacing is considered. A clear downside of the Barata-Fredenhagen approach is that, while the scattering amplitude is expected to have a continuum limit, the same can not be said for the approximant of the scattering amplitude, i.e. the quantity which corresponds to $\mathcal{S}_c(\sigma,\epsilon)$ in their work.

The functions $\Upsilon(\tau n;\vec{p})$ increase exponentially with $n$, see eq.~\eqref{eq:euclidean:Upsilon}, meaning that the lattice correlators at larger-time separations are multiplied with exponentially larger weights in eq.~\eqref{eq:remark}. This feature captures the intuitive idea that scattering amplitudes are encoded, in a highly non-trivial fashion, in the long-distance behavior of the Euclidean correlators. Even though one can design different approximation strategies for $\mathcal{S}_c(\sigma,\epsilon)$, we believe that this feature is quite general: in essence, this is the way in which the inverse Laplace transform of the Euclidean correlator is reconstructed. It is important to notice that, for a given target accuracy, only a finite number of Euclidean time separations are needed. This number will increase as $\epsilon$ and $\sigma$ are decreased. The statistical noise of the correlator (which for generic lattice QCD correlators increases exponentially at large euclidean-time separations) limits the accuracy that can be achieved.

The accuracy of the approximation can be increased at the expenses of the statistical precision and vice versa. In practice, one may want to design an optimization strategy which minimizes the total error for a given set of numerically-determined Euclidean correlation functions. To do this one can consider a straightforward generalization of the  HLT method  in which the so-called error-functional,
\begin{gather}
B[w]=\sum_{n,n^\prime, b,b^\prime} w^{\sigma,\epsilon}_{n,b}\ \mathrm{Cov}(n,b;n^\prime, b^\prime)\, w^{\sigma,\epsilon}_{n^\prime,b^\prime} \ ,
\end{gather}
where $ \mathrm{Cov}(n,b;n^\prime, b^\prime)$ is the statistical covariance-matrix of the smeared correlator $\mathcal{C}(n,b)$,
is added to the left-hand side of eq.~\eqref{eq:approx:P}, that we now call $A[w]$ and that in the language of ref.~\cite{Hansen:2019idp} corresponds to the norm-functional. More explicitly, accuracy and statistical precision can be balanced by obtaining the  coefficients $w^{\sigma,\epsilon}_{n,b}$ from the minimization of the following linear combination
\begin{gather}
A[w]+\lambda B[w]
\end{gather}
of the norm and error functionals and then by studying the stability of the resulting approximation of the scattering amplitude upon variations of the algorithmic parameter $\lambda$ (see ref.~\cite{Bulava:2021fre} for more details).

Finally, we want to stress that throughout this paper we work with smooth wave packets, i.e. with asymptotic states characterized by smooth normalizable wave functions. From the mathematical perspective, this is essential in order to derive our results within the rigorous axiomatic framework. From the numerical perspective, this is also essential in order to have an estimator with exponentially suppressed finite-volume corrections. Often one is interested in the connected amplitude $S_c(\vec{p})$ corresponding to the scattering of particles in plane-wave states with momenta $\vec{p}_A$. The plane-wave scattering amplitude is proportional to the delta of energy-momentum conservation, i.e.
\begin{gather}
S_c(\vec{p})= (2\pi)^4\delta\Bigg(\sum_{A=1}^{M+N} \eta_A E(\vec{p}_A) \Bigg)\delta^3\Bigg( \sum_{A=1}^{M+N} \eta_A \vec{p}_A \Bigg) T_c(\vec{p})\ .
\end{gather}
Given some momenta $\bar{\vec{p}}_A$ satisfying the energy-momentum conservation, assuming that $T_c(\bar{\vec{p}})$ is a continuous function of its arguments, it can be obtained by means of the following limiting procedure:
\begin{gather}
T_c(\bar{\vec{p}}) = \lim_{\sigma_\vec{p} \to 0} \frac{S_c(\check{f}_{\bar{\vec{p}},\sigma_\vec{p}})}{N(\bar{\vec{p}},\sigma_{\vec{p}})}
\ ,
\end{gather}
where $S_c(\check{f}_{\bar{\vec{p}},\sigma_\vec{p}})$ is the scattering amplitude for particles with smooth wave functions $\check{f}_{\bar{\vec{p}}_A,\sigma_\vec{p}}(\vec{p}_A)$ centered in $\bar{\vec{p}}_A$ and with width proportional to the parameter $\sigma_{\vec{p}}$ (see appendix~\ref{app:compact} for a concrete example). The normalization factor is given by
\begin{gather}
N(\bar{\vec{p}},\sigma_{\vec{p}}) = \int \Bigg[ \prod_{A=1}^{M+N} \frac{d^3 \vec{p}_A}{(2\pi)^3} \check{f}_{\bar{\vec{p}}_A,\sigma_\vec{p}}^{(*)}(\vec{p}_A) \Bigg]
(2\pi)^4\delta\Bigg(\sum_{A=1}^{M+N} \eta_A E(\vec{p}_A) \Bigg) \delta^3\Bigg( \sum_{A=1}^{M+N} \eta_A \vec{p}_A \Bigg)
\ .
\end{gather}

In summary, the proposed method to extract scattering amplitudes from Euclidean correlators is mathematically robust but its numerical applicability with lattice data may by limited by the level of precision and accuracy that can be presently reached. A detailed numerical investigation is needed to assess this important point. However, the numerical accuracy and precision of lattice simulations is systematically improvable, and we have little doubts that our approach will become useful in the future.

%--------------------------------------------------------------
\begin{acknowledgement}%
% !TEX root = paper.tex

N.T. thanks Massimo Testa for opening his eyes on the beauties of Haag-Ruelle scattering theory. We thank John Bulava for useful comments on the manuscript. We warmly thank O. and T. for their constant, patient and silent support at all stages of this work. N.T. is supported by the Italian Ministry of University and Research (MUR) under the grant PNRR-M4C2-I1.1-PRIN 2022-PE2 Non-perturbative aspects of fundamental interactions, in the Standard Model and beyond F53D23001480006 funded by E.U. - NextGenerationEU. A.P. thanks Alberto Ramos for generously hosting him for a period of three months at IFIC in Valencia, where this work was finalized.

\end{acknowledgement}

\clearpage
\appendix

\section{Smooth functions with compact support}\label{app:compact}
The Haag-Ruelle construction and, hence, our analysis heavily rely on the existence of smooth functions with compact support. In this brief appendix, we want to provide some concrete examples for the reader who may not be familiar with these mathematical objects. A classical example is the so-called bump function, i.e.
\begin{gather}
B(x)=
\begin{cases}
\exp\left( -\frac{x^2}{1-x^2}\right) \qquad &\mathrm{for } x^2< 1
\\
0 &\mathrm{for } x^2\ge1
\end{cases}
\ .
\end{gather}
The closed support of $B(x)$ is the interval $[-1,1]$ and it is easy to prove that this function is infinitely differentiable everywhere.
Moreover, its normalization has been chosen in such a way that $B(0)=1$. We point out that the bump function, like any other smooth function with compact support, is Schwartz. A smooth function with compact support in $[a,b]$ is simply obtained by considering $B((2x-a-b)/(b-a))$.

An explicit example of smooth wave function $\check{f}(\vec p)$ with compact support centered around the momentum $\bar{\vec{p}}$ can easily be written int terms of the bump function:
\begin{gather}
\check{f}(\vec p) = B\left( \frac{\| \vec p-\bar{\vec{p}} \|}{\sigma_{\vec{p}}}\right)\, \frac{(2\pi)^3}{\sigma_{\vec{p}}^3\int d^3\vec k\, B(\| \vec k \|)}\ ,
\end{gather}
where $\sigma_{\vec{p}}$ is a numerical constant that regulates the width of the wave function. In particular, the wave function vanishes for $ \|  \vec p-\bar{\vec{p}} \| \ge  \sigma_{\vec{p}}$. The normalization has been chosen in such a way that
\begin{gather}
\lim_{\sigma_{\vec{p}}\to 0} \check{f}(\vec p) = (2\pi)^3 \delta^3(\vec p-\bar{\vec{p}})\ .
\end{gather}

With this choice of momentum wave function, an explicit example for the function $\zeta(\omega)$ which appears in the construction of Haag's operators (see eq.~\eqref{eq:intro:f}) can be obtained by following the construction described in appendix~\ref{app:haag} and by using the bump function again. For instance, one can choose
\begin{gather}
\zeta(\omega) = B\left( \frac{2\omega-\omega_1-\omega_2}{\omega_2-\omega_1}\right)\ ,
\end{gather}
with the definitions
\begin{gather}
\Lambda= \| \bar{\vec{p}} \| + \sigma_{\vec{p}}
\, \\
\omega_1 = \frac{3\Lambda - 4\sqrt{m^2+\Lambda^2} + \sqrt{4m^2+\Lambda^2}}{4}
\ , \\
\omega_2 = \frac{3\sqrt{4m^2+\Lambda^2} - 4\sqrt{m^2+\Lambda^2} + \Lambda}{4}
\ .
\end{gather}

An explicit example for the function $\Phi(\tau)$ appearing in the Haag-Ruelle kernel (see eq.~\eqref{eq:tantalo:T-def} and eq.~\eqref{eq:spectral:K}) is given by
\begin{gather}
\bar \Phi(\tau)
=
\begin{cases}
\frac{1}{K_1(1)} \exp\left( -\frac{1}{2(\tau-\bar \tau)} -\frac{\tau-\bar \tau}{2}\right)
\qquad & \mathrm{for } \tau>\bar \tau
\\
0 & \mathrm{for } \tau \le \bar \tau
\end{cases}
\ ,
\end{gather}
where $\bar{\tau}$ is and arbitrary positive number and $K_n(z)$ denotes the modified Bessel function of the second kind. The function $\Phi(\tau)$ has support in $[\bar \tau,+\infty)$, is Schwartz and has unit integral. With this choice, the Fourier transform of $\Phi(\tau)$ has a closed expression in terms of the Bessel function $K_1(z)$, i.e.
\begin{gather}
\tilde \Phi(\omega) = e^{-i\bar \tau \omega} \frac{K_1(\sqrt{1-2i \omega})}{K_1(1) \sqrt{1-2i \omega}}\ .
\label{eq:phitildeexplicit}
\end{gather}

Finally, we provide an explicit example for the function $\tilde h(\omega)$ which appears in our approximation formulae (see eq.~\eqref{eq:tantalo:T-def} and eq.~\eqref{eq:approx:S-sigma-eps}). Given $\bar \Delta> 0$ (see eq.~\eqref{eq:approx:Kbb}), one can simply set
\begin{gather}
\tilde h(\omega) = B\left( \frac{\omega}{\bar \Delta}\right)\ .
\label{eq:htildeexplicit}
\end{gather}

\section{Construction of Haag operators}\label{app:haag}
\textbf{Statement.} Given a function $\check{f}(\vec{p})$ with compact support, a smooth function $\zeta(\omega)$ exists such that: \textit{(a)} $0 \le \zeta(\omega) \le 1$, \textit{(b)}  $\zeta(0)=1$, \textit{(c)} the function $p \mapsto \zeta(p_0 {-} E(\vec{p})) \check{f}(\vec{p})$ has closed support inside $|\vec{p}| < p_0 < \sqrt{4m^2+\vec{p}^2}$.

\bigskip

\textbf{Proof.} Given $\mu_2 \ge \mu_1 \ge 0$, the function $g(\lambda^2) = \sqrt{\mu_2^2 + \lambda^2} - \sqrt{\mu_1^2 + \lambda^2}$ is monotonously decreasing for $\lambda^2 \ge 0$. This is easily checked by rewriting
\begin{gather}
    g(\lambda^2) = \frac{ \mu_2^2 - \mu_1^2 }{ \sqrt{\mu_2^2 + \lambda^2} + \sqrt{\mu_1^2 + \lambda^2} } \ ,
\end{gather}
and noticing that the numerator is positive and the denominator is monotonously increasing in $\lambda^2$.

Define the largest momentum allowed by the considered wave function, more precisely:
\begin{gather}
    \Lambda = \sup \{ \| \vec{p} \| \text{ s.t. } \vec{p} \in \supp \check{f} \} \ ,
\end{gather}
which is finite since $\check{f}(\vec{p})$ has compact support. Choose some $\epsilon>0$ and define
\begin{gather}
    \omega_1 = \Lambda - \sqrt{m^2+\Lambda^2} + \epsilon
    \ , \\
    \omega_2 = \sqrt{4m^2+\Lambda^2} - \sqrt{m^2+\Lambda^2} - \epsilon
    \ .
\end{gather}
We require that $\epsilon$ is small enough such that $\omega_1 < \omega_2$. Then we can choose $\zeta(\omega)$ to be a smooth function with values in the interval $[0,1]$, which is equal to one for $\omega=0$ and vanishes for $\omega \not\in [\omega_1,\omega_2]$. Assume now that $p$ is in the support of $\zeta(p_0 {-} E(\vec{p})) \, \check{f}(\vec{p})$. Then $\|\vec{p}\| \le \Lambda$ and we can use the above observation concerning the function $g(\lambda^2)$ twice (the first time with $\mu_2=2m$ and $\mu_1=m$, and the second time with $\mu_2=m$ and $\mu_1=0$) to derive the following inequalities:
\begin{gather}
    p_0 - E(\vec{p}) \le \omega_2 = \sqrt{4m^2+\Lambda^2} - \sqrt{m^2+\Lambda^2} - \epsilon
    \le \sqrt{4m^2+\vec{p}^2} - \sqrt{m^2+\vec{p}^2} - \epsilon
    \ , \\
    p_0 - E(\vec{p}) \ge \omega_1 = - \left( \sqrt{m^2+\Lambda^2} - \Lambda \right) + \epsilon
    \ge - \left( \sqrt{m^2+\vec{p}^2} - |\vec{p}| \right) + \epsilon
    \ ,
\end{gather}
i.e. $|\vec{p}| + \epsilon \le p_0 \le \sqrt{4m^2+\vec{p}^2} - \epsilon$.

\section{Asympotic behaviour of \texorpdfstring{$\mathcal{S}_c(\sigma)$}{Sc(σ)}}\label{app:transition}
\textbf{Statement.} In this appendix we use the same notation and definitions given in section~\ref{subsec:tantalo}. For $\sigma>0$ we define the integrated transition probability
\begin{gather}
    \mathcal{S}(\sigma) = \sigma \int dt \, ds \, \Phi(t\sigma) \, h(s) \, \Big\langle \Psi_\sout\left( \tfrac{t}{2}{-}s \right) \Big| \Psi_\sin\left( -\tfrac{t}{2}{-}s \right) \Big\rangle
    \label{eq:app:transition:T-def}
\end{gather}
and the scattering amplitude
\begin{gather}
    S = \langle \Psi_\sout(+\infty) | \Psi_\sin(-\infty) \rangle
    \ .
\end{gather}
Then, the following equality holds:
\begin{gather}
    \mathcal{S}(\sigma)
    \overset{\sigma \to 0^+}{=}
    S
    + O(\sigma^{r})
    \ ,
    \label{eq:app:transition:TS}
\end{gather}
where $r$ can be an arbitrary positive number in the case of non-overlapping velocities and $r=1/2$ in the general case.

\bigskip

\textbf{Proof.} We start from the following identity, obtained by using the fact that $\Phi(\tau)$ and $h(s)$ have unit integrals:
\begin{gather}
    \label{eq:app:tr:diff-1}
    \mathcal{S}(\sigma) - S
    =
    \mathcal{S}(\sigma) - S \int d\tau \, ds \, \Phi(\tau) \, h(s)
    \\ \nonumber =
    \int d\tau \, ds \, \Phi(\tau) \, h(s) \, \Big\langle \Psi_\sout\left( \tfrac{\tau}{2\sigma}{-}s \right) - \Psi_\sout(+\infty) \Big| \Psi_\sin\left( -\tfrac{\tau}{2\sigma}{-}s \right) \Big\rangle
    \\ \nonumber \phantom{ = }
    + \int d\tau \, ds \, \Phi(\tau) \, h(s) \,  \Big\langle \Psi_\sout(+\infty) \Big| \Psi_\sin\left( -\tfrac{\tau}{2\sigma}{-}s \right) - \Psi_\sin(-\infty) \Big\rangle
    \ .
\end{gather}
The triangular inequality and Cauchy-Schwartz inequality imply
\begin{gather}
    \label{eq:app:tr:diff-2}
    \left| \mathcal{S}(\sigma) - S \right|
    \\ \nonumber \le
    \int d\tau \, ds \, |\Phi(\tau) h(s)| \, \Big\| \Psi_\sout\left( \tfrac{\tau}{2\sigma}{-}s \right) - \Psi_\sout(+\infty) \Big\| \, \Big\| \Psi_\sin\left( -\tfrac{\tau}{2\sigma}{-}s \right) \Big\|
    \\ \nonumber \phantom{ \le }
    + \int d\tau \, ds \, |\Phi(\tau) h(s)| \, \Big\| \Psi_\sout(+\infty) \Big\| \, \Big\| \Psi_\sin\left( -\tfrac{\tau}{2\sigma}{-}s \right) - \Psi_\sin(-\infty) \Big\|
    \\ \nonumber \le
    \| \Psi_\sin \|_\infty \int d\tau \, ds \, |\Phi(\tau) h(s)| \, \Big\| \Psi_\sout\left( \tfrac{\tau}{2\sigma}{-}s \right) - \Psi_\sout(+\infty) \Big\|
    \\ \nonumber \phantom{ \le }
    + \| \Psi_\sout \|_\infty \int d\tau \, ds \, |\Phi(\tau) h(s)| \, \Big\| \Psi_\sin\left( -\tfrac{\tau}{2\sigma}{-}s \right) - \Psi_\sin(-\infty) \Big\|
    \ .
\end{gather}
Here we have also used the fact that $\Psi_\as(t)$ (for $\as = \sin, \sout$) is infinitely differentiable in $t$ and has a finite limits for $t \to \pm \infty$, i.e.
\begin{gather}
    \| \Psi_\as \|_\infty = \sup_t \| \Psi_\as(t) \| < +\infty \ .
\end{gather}
Since the support of $\Phi(\tau)$ is contained in $[\bar{\tau},+\infty)$ for some $\bar{\tau}>0$ (together with an $s \to -s$ substitution in the first integral), we obtain:
\begin{gather}
    \label{eq:app:tr:diff-3}
    \left| \mathcal{S}(\sigma) - S \right|
    \\ \nonumber \le
    \| \Psi_\sin \|_\infty \int_{\bar{\tau}}^{+\infty} d\tau \int_{-\infty}^\infty ds \, |\Phi(\tau) h(-s)| \, \Big\| \Psi_\sout\left( \tfrac{\tau}{2\sigma}{+}s \right) - \Psi_\sout(+\infty) \Big\|
    \\ \nonumber \phantom{ \le }
    + \| \Psi_\sout \|_\infty \int_{\bar{\tau}}^{+\infty} d\tau \int_{-\infty}^\infty ds \, |\Phi(\tau) h(s)| \, \Big\| \Psi_\sin\left( -\tfrac{\tau}{2\sigma}{-}s \right) - \Psi_\sin(-\infty) \Big\|
    \ .
\end{gather}

Let us focus on the norms appearing in the above integrals. We use the fact that the states $\Psi_\as(t)$ are infinitely differentiable and reach their asymptotic values for $t \to \pm \infty$ with an error that decreases like $|t|^{-r}$, where $r$ can be an arbitrary positive number in the case of non-overlapping velocities and $r=1/2$ in the general case. In particular
\begin{gather}
    \| \Psi_\sout( t ) - \Psi_\sout(+\infty) \| \le
    \begin{cases}
    \| \Psi_\sout( t ) \| + \| \Psi_\sout(+\infty) \| \le 2 \| \Psi_\sout \|_\infty \qquad & \text{for } t<0 \\[4pt]
    C^\sout_r (1+t^2)^{-\frac{r}{2}} \qquad & \text{for } t \ge 0
    \end{cases}
    \ .
\end{gather}
In writing this inequality we have assumed an arbitrary unit system. Units can be restored by replacing $(1+t^2)$ in the above expression with $(L^2+t^2)$ where $L$ is an arbitrary length scale. Calculating this for $t = \tfrac{\tau}{2\sigma} + s > 0$, and using the inequality
\begin{gather}
    \frac{\beta^2}{1+(\alpha+\beta)^2} \le \sup_{\beta} \frac{\beta^2}{1+(\alpha+\beta)^2} = 1+\alpha^2 \ ,
\end{gather}
one gets
\begin{gather}
    \| \Psi_\sout\left( \tfrac{\tau}{2\sigma} {+} s \right) - \Psi_\sout(+\infty) \| \le
    \begin{cases}
    2 \| \Psi_\sout \|_\infty \qquad & \text{for } s < -\tfrac{\tau}{2\sigma} \\[4pt]
    C^\sout_r (1+s^2)^\frac{r}{2} \left(\frac{\tau}{2\sigma}\right)^{-r}  \qquad & \text{for } s \ge -\tfrac{\tau}{2\sigma}
    \end{cases}
    \ .
\end{gather}
Repeating the argument for the incoming state, one gets:
\begin{gather}
    \| \Psi_\sin\left( - \tfrac{\tau}{2\sigma} {-} s \right) - \Psi_\sin(-\infty) \| \le
    \begin{cases}
    2 \| \Psi_\sin \|_\infty \qquad & \text{for } s < -\tfrac{\tau}{2\sigma} \\[4pt]
    C^\sin_r (1+s^2)^\frac{r}{2} \left(\frac{\tau}{2\sigma}\right)^{-r}  \qquad & \text{for } s \ge -\tfrac{\tau}{2\sigma}
    \end{cases}
    \ .
\end{gather}

In combination with eq.~\eqref{eq:app:tr:diff-3}, the above estimates yield
\begin{gather}
    \label{eq:app:tr:diff-4}
    \left| \mathcal{S}(\sigma) - S \right|
    \\ \nonumber \le
    2 \| \Psi_\sin \|_\infty \| \Psi_\sout \|_\infty \int_{\bar{\tau}}^{+\infty} d\tau \, |\Phi(\tau)| \int_{-\infty}^{-\frac{\tau}{2\sigma}} ds \,  \{ |h(-s)| + |h(s)| \}
    \\ \nonumber \phantom{ \le }
    + 2^r C^\sout_r \sigma^r \| \Psi_\sin \|_\infty \int_{\bar{\tau}}^{+\infty} d\tau \, |\Phi(\tau)| \, \tau^{-r} \int_{-\frac{\tau}{2\sigma}}^\infty ds \, |h(-s)| \, (1+s^2)^\frac{r}{2}
    \\ \nonumber \phantom{ \le }
    + 2^r C^\sin_r \sigma^r \| \Psi_\sout \|_\infty \int_{\bar{\tau}}^{+\infty} d\tau \, |\Phi(\tau)| \, \tau^{-r} \int_{-\frac{\tau}{2\sigma}}^\infty ds \, |h(s)| \, (1+s^2)^\frac{r}{2}
    \ .
\end{gather}
The second and third integrals over $s$ can be extended to the whole real axis. In the first integral over $s$, we use that fact that $h(s)$ is Schwartz, which implies that a constant $D_r$ exists such that
\begin{gather}
    |h(s)| \le D_r |s|^{-r-1}
\end{gather}
for every $s \neq 0$ and, therefore,
\begin{gather}
    \int_{-\infty}^{-\frac{\tau}{2\sigma}} ds \,  \{ |h(-s)| + |h(s)| \}
    \le
    2 D_r \int_{-\infty}^{-\frac{\tau}{2\sigma}} ds \, s^{-r-1}
    =
    \frac{2^{1+r} D_r}{r} \tau^{-r} \sigma^r
    \ .
\end{gather}
Plugging this back in eq.~\eqref{eq:app:tr:diff-4}, we finally get
\begin{gather}
    \sigma^{-r} \left| \mathcal{S}(\sigma) - S \right|
    \\ \nonumber \le
    \frac{2^{2+r} D_r}{r} \| \Psi_\sin \|_\infty \| \Psi_\sout \|_\infty \int_{\bar{\tau}}^{+\infty} d\tau \, \tau^{-r} |\Phi(\tau)|
    \\ \nonumber \phantom{ \le }
    + 2^r C^\sout_r \| \Psi_\sin \|_\infty \int_{\bar{\tau}}^{+\infty} d\tau \, |\Phi(\tau)| \, \tau^{-r} \int_{-\infty}^\infty ds \, |h(-s)| \, (1+s^2)^\frac{r}{2}
    \\ \nonumber \phantom{ \le }
    + 2^r C^\sin_r \| \Psi_\sout \|_\infty \int_{\bar{\tau}}^{+\infty} d\tau \, |\Phi(\tau)| \, \tau^{-r} \int_{-\infty}^\infty ds \, |h(s)| \, (1+s^2)^\frac{r}{2}
    \ ,
\end{gather}
where the right-hand side does not depend on $\sigma$ and is finite because $\Phi(\tau)$ and $h(s)$ are Schwartz. This concludes the proof.

\section{Estimates for \texorpdfstring{$\mathfrak{N}_\omega$}{Nω} and \texorpdfstring{$\mathfrak{N}_{\vec{p}}$}{Np}}\label{app:order}
\textbf{Statement.} Assume that, for every positive integer $q$, a tempered distribution $C_{q}$ exists such that
\begin{gather}
    \left| W^g_c(x) \right| \le C_{q}(g) \frac{ [ 1 + m \, d_t(x_0) ]^q }{ [ 1 + m \, d_s(\vec{x}) ]^{q-1} }
    \label{eq:app-order:BFbound}
\end{gather}
where $d_s(\vec{x})$ is the diameter of the set of spatial points, i.e.
\begin{gather}
    d_s(\vec{x}) = \max_{A,B=1,\dots,M} \| \vec{x}_A - \vec{x}_B \|_2 \ ,
\end{gather}
and $d_t(x_0)$ is the maximum time separation between consecutive operators as they appear in the right-hand side of eq.~\eqref{eq:spec:Wgc(x)},. i.e.
\begin{gather}
    d_t(x_0) = \max_{A=1,\dots,M+N-1} | \tau_A |
    \ ,
\end{gather}
with the definitions:
\begin{gather}
    \label{eq:app-order:tau}
    \tau_{A<M} = x_{A,0}-x_{A+1,0}
    \ , \quad
    \tau_{A>M} = x_{A+1,0}-x_{A,0}
    \ , \quad
    \tau_M = x_{M,0}-x_{M+N,0}
    \ .
\end{gather}
Then, a representation of the type~\eqref{eq:order:rho-R} exists with
\begin{gather}
    \label{eq:app-order:N}
    \mathfrak{N}_\omega = 2(M+N) \ , \quad \mathfrak{N}_{\vec{p}} = 0 \ .
\end{gather}

\bigskip

\textbf{Proof.} We introduce a smooth real function $\tilde{u}(p)$ with $p \in \mathbb{R}^4$ which satisfies the following properties:
\begin{gather}
    \tilde{u}(p)
    =
    \begin{cases}
        1 \quad & \text{if } \|p\|_\infty \le \frac{1}{2} \\[4pt]
        0 \quad & \text{if } \|p\|_\infty \ge 1
    \end{cases}
    \ , \qquad
    \sum_{n \in \mathbb{Z}^4} \tilde{u}(p-n) = 1
    \ .
\end{gather}
Notice that the infinite sum over $n$ does not present issues of convergence since, for every $p$, only a finite number of terms is different from zero. Such a function can be constructed quite explicitly and we will not dwell on its existence. Given some $(n)=(n_1,\dots,n_{M+N-1}) \in \mathbb{Z}^{4(M+N-1)}$, we define
\begin{gather}
    \tilde{u}_{n_A}(p_A) = \tilde{u}(p_A-n_A)
    \ , \\
    \tilde{v}_n(p_{M+N}) = \tilde{u}\left( \frac{p_A}{r_n} \right)
    \qquad \text{with } r_n = 2 \sum_{A=1}^{M+N-1} \left( 1 + \| n_A \|_\infty \right)
    \ .
\end{gather}
In coordinate space, these relations read
\begin{gather}
    u_{n_A}(x_A)
    = e^{-in_Ax_A} u(x_A)
    \ , \\
    v_n(x_{M+N})
    = r_n^4 u\left( r_n x_{M+N} \right)
    \ .
\end{gather}
The function $\tilde{v}_n$ is designed in such a way that
\begin{gather}
    \label{eq:app-order:vtilde-times-utilde}
    \tilde{v}_n\left( - \!\! \sum_{A=1}^{M+N-1} \eta_A p_A \right)
    \Bigg[ \prod_{A=1}^{M+N-1} \tilde{u}_{n_A}(p_A) \Bigg]
    =
    \Bigg[ \prod_{A=1}^{M+N-1} \tilde{u}_{n_A}(p_A) \Bigg]
    \ .
\end{gather}

Let $W^n_c(x)$ be the smeared connected Wightman function defined as in eq.~\eqref{eq:spec:Wgc(x)} with the choice $g_A(x) = u_{n_A}(x)$ for $A=1,\dots,M+N-1$ and $g_{M+N}(x) = v_n(x)$. Eq.~\eqref{eq:spectral:rho} implies
\begin{gather}
    \label{eq:app-order:Wnc-rho-1}
    \int \Bigg[ \prod_{A=1}^{M+N-1} d^4x_A \, e^{i \eta_A p_Ax_A} \Bigg] \, W^n_c(x_1,\dots,x_{M+N-1},0)
    \\ \nonumber =
    \Bigg[ \prod_{A=1}^{M+N-1} \tilde{u}(p_A-n_A) \Bigg] \, \rho_c(\omega,\vec{p})
    \ ,
\end{gather}
where we have used eq.~\eqref{eq:app-order:vtilde-times-utilde} to remove the function $\tilde{v}_n$ in the right-hand side. Noticing that $\tilde{u}\left(\frac{p_A-n_A}{2}\right) \tilde{u}(p_A-n_A) = \tilde{u}(p_A-n_A)$, we obtain the following representation for the spectral density:
\begin{gather}
\label{eq:app-order:rho}
\rho_c(\omega,\vec{p})
=
\sum_{n_1,\dots,n_{M+N-1} \in \mathbb{Z}^4}
\Bigg[ \prod_{A=1}^{M+N-1} \tilde{u}(p_A-n_A) \Bigg] \rho_c(\omega,\vec{p})
\\
\nonumber
=
\sum_{n_1,\dots,n_{M+N-1} \in \mathbb{Z}^4}
\Bigg[ \prod_{A=1}^{M+N-1} \tilde{u}\left(\frac{p_A-n_A}{2}\right) \tilde{u}(p_A-n_A) \Bigg] \rho_c(\omega,\vec{p})
\\
\nonumber
= \sum_{n_1,\dots,n_{M+N-1} \in \mathbb{Z}^4}
\Bigg[ \prod_{A=1}^{M+N-1} \tilde{u}\left(\frac{p_A-n_A}{2}\right) \Bigg]
\\
\nonumber \phantom{= \qquad }
\times
\int \Bigg[ \prod_{A=1}^{M+N-1} d^4x_A \, e^{i \eta_A p_Ax_A} \Bigg] \, W^n_c(x_1,\dots,x_{M+N-1},0)
\ .
\end{gather}

Let us focus on the integral in the above expression. We define the function
\begin{gather}
    F_n(\tau,\vec{x})
    =
    W^n_c(x_1,\dots,x_{M+N-1},0)
\end{gather}
where the variables $\tau_A$ are related to $x_{A,0}$ by eq.~\eqref{eq:app-order:tau} with the constraint $x_{M+N}=0$. The relation between $\tau_A$ and $x_{A,0}$ is easily inverted (assuming again $x_{M+N}=0$):
\begin{gather}
    x_{A,0}
    =
    \begin{cases}
    \sum_{B=A}^{M} \tau_B \quad & \text{if } A \le M \\[4pt]
    - \sum_{B=A}^{M+N-1} \tau_B \quad & \text{if } A > M
    \end{cases}
    \ .
\end{gather}
A quick calculation shows that
\begin{gather}
    \sum_{A=1}^{M+N-1} \eta_A p_A x_A = \sum_{A=1}^{M+N-1} \left\{ - q_{A,0} \tau_A - \eta_A \vec{p}_A \vec{x}_A \right\}
    \ ,
\end{gather}
where the variables $q_{A,0}$ are defined by eq.~\eqref{eq:spectral:q}. Consider the integral in the last line of eq.~\eqref{eq:app-order:rho}. By substituting then integration variable $x_{A,0}$ with $\tau_A$, the Jacobian determinant being one, one obtains
\begin{gather}
    \label{eq:app-order:Wtilde-Ftilde}
    \int \Bigg[ \prod_{A=1}^{M+N-1} d^4x_A \, e^{i \eta_A p_Ax_A} \Bigg] \, W^n_c(x_1,\dots,x_{M+N-1},0)
    \\ \nonumber
    \phantom{\int \Bigg[ \prod_{A=1}^{M+N-1} d^4x_A \, e^{i \eta_A p_Ax_A} \Bigg]}
    =\int \Bigg[ \prod_{A=1}^{M+N-1} d\tau_A d^3\vec{x}_A \, e^{-i q_{A,0} \tau_A -i \eta_A \vec{p}_A\vec{x}_A} \Bigg] \, F_n(\tau,\vec{x})
    \ .
\end{gather}

The following elementary inequalities (assuming again $x_{M+N}=0$):
\begin{gather}
    d_s(\vec{x}) \ge \max_{A=1,\dots,M+N-1} \|\vec{x}_A\|_2 \ge \max_{A=1,\dots,M+N-1} \max_{k=1,2,3} |x_{A,k}| = \| \vec{x} \|_\infty
    \ , \\
    1 + m \, \| \tau \|_\infty \le \sqrt{2} \sqrt{ 1 + m^2 \| \tau \|_\infty^2 }
    \ ,
\end{gather}
together with the assumption~\eqref{eq:app-order:BFbound}, imply the bound
\begin{gather}
    \label{eq:app-order:bound-F-1}
    \left| F_n(\tau,\vec{x}) \right| \le 2^{\frac{q}{2}} C_{n,q} \frac{ ( 1 + m^2 \| \tau \|_\infty^2 )^{\frac{q}{2}} }{ ( 1 + m \| \vec{x} \|_\infty )^{q-1} }
    \ ,
\end{gather}
where
\begin{gather}
    \label{eq:app-order:Cn}
    C_{n,q} = C_q( u_{n_1} \otimes \cdots \otimes u_{n_{M+N-1}} \otimes v_n ) \ .
\end{gather}
We introduce the following family of functions labeled by $A=1,\dots,M+N-1$:
\begin{gather}
    \label{eq:app-order:FA}
    F_{n,A}(\tau,\vec{x})
    =
    \begin{cases}
        F_n(\tau,\vec{x}) \quad & \text{if } |\tau_A| = \| \tau \|_\infty \\[4pt]
        0 \quad & \text{otherwise}
    \end{cases}
    \ ,
\end{gather}
which satisfy the identity
\begin{gather}
    \label{eq:app-order:sum-F}
    F_n(\tau,\vec{x}) = \sum_{A=1}^{M+N-1} F_{n,A}(\tau,\vec{x})
\end{gather}
valid almost everywhere. Since $q$ is an arbitrary non-negative number, we choose $q = 3(M+N)/2$. Eq.~\eqref{eq:app-order:bound-F-1} implies the bound
\begin{gather}
    \label{eq:app-order:bound-F-2}
    \left| F_{n,A}(\tau,\vec{x}) \right|
    \le
    2^{\frac{3(M+N)}{4}} C_{n,\frac{3(M+N)}{2}} \frac{ | 1 -i m \, \tau_A |^{\frac{3(M+N)}{2}} }{ ( 1 + m \| \vec{x} \|_\infty )^{\frac{3(M+N-1)+1}{2}} }
    \ .
\end{gather}
Then, the function
\begin{gather}
    \label{eq:app-order:G}
    G_{n,A}(\tau,\vec{x})
    =
    ( 1 - i m \tau_A )^{-2(M+N)}
    F_{n,A}(\tau,\vec{x})
\end{gather}
is square integrable, and so is its Fourier transform
\begin{gather}
    \tilde{G}_{n,A}(q_0,\vec{p})
    =
    \int \Bigg[ \prod_{B=1}^{M+N-1} d\tau_B d^3\vec{x}_B \, e^{-i q_{B,0} \tau_B -i \eta_B \vec{p}_B\vec{x}_B} \Bigg] \, G_{n,A}(\tau,\vec{x})
    \ .
\end{gather}
Eqs.~\eqref{eq:app-order:Wtilde-Ftilde}, \eqref{eq:app-order:sum-F} and \eqref{eq:app-order:G} yield
\begin{gather}
    \label{eq:app-order:Wtilde-Ftilde-2}
    \int \Bigg[ \prod_{A=1}^{M+N-1} d^4x_A \, e^{i \eta_A p_Ax_A} \Bigg] \, W^n_c(x_1,\dots,x_{M+N-1},0)
    \\ \nonumber =
    \sum_{A=1}^{M+N-1} \tilde{F}_{n,A}(q_0,\vec{p})
    \\ \nonumber =
    \sum_{A=1}^{M+N-1} \left( 1 + m \frac{\partial}{\partial q_{A,0}} \right)^{2(M+N)}
    \tilde{G}_{n,A}(q_0,\vec{p})
    \\ \nonumber =
    \sum_{A=1}^{M+N-1} \left( 1 + m \frac{\partial}{\partial \omega_A} \right)^{2(M+N)}
    \mathring{G}_{n,A}(\omega,\vec{p})
    \ ,
\end{gather}
where the derivatives must be interpreted in the distributional sense. In the last step, we have used the definition
\begin{gather}
    \label{eq:app-order:Gring}
    \mathring{G}_{n,A}(\omega,\vec{p}) = \tilde{G}_{n,A}(q_0,\vec{p})
\end{gather}
and the relation between $\omega_A$ and $q_{A,0}$, which can be worked out explicitly using eqs.~\eqref{eq:spectral:q} and \eqref{eq:spectral:omega}:
\begin{gather}
    \omega_A =
    \begin{cases}
        q_{B,0} - \sum_{B=1}^A E(\vec{p}_B)
        \quad & \text{for } 1 \le A \le M \\[4pt]
        q_{B,0} - \sum_{B=M+1}^A E(\vec{p}_B)
        \quad & \text{for } M+1 \le A \le M+N-1
    \end{cases}
    \ .
\end{gather}
Notice that the function $\mathring{G}_{n,A}(\omega,\vec{p})$ is also square integrable. Combining eqs.~\eqref{eq:app-order:rho} and \eqref{eq:app-order:Wtilde-Ftilde-2}, we obtain the following representation for the spectral density:
\begin{gather}
    \label{eq:app-order:rho-G}
    \rho_c(\omega,\vec{p})
    =
    \sum_{n_1,\dots,n_{M+N-1} \in \mathbb{Z}^4}
    \sum_{A=1}^{M+N-1}
    w_n(\omega,\vec{p}) \left( 1 + m \frac{\partial}{\partial \omega_A} \right)^{2(M+N)}
    \mathring{G}_{n,A}(\omega,\vec{p})
    \ ,
\end{gather}
with the definition
\begin{gather}
    \label{eq:app-order:wn}
    w_n(\omega,\vec{p}) = \prod_{A=1}^{M+N-1} \tilde{u}\left(\frac{p_A-n_A}{2}\right) \ .
\end{gather}
One checks that $w_n(\omega,\vec{p})$ is smooth with compact support. Moreover, $w_n$ restricted to any compact subset of $\mathbb{R}^{4(M+N-1)}$ is not identically zero only for a finite number of values of $(n_1,\dots,n_{M+N-1})$. Some lengthy but straightforward algebra yields:
\begin{gather}
    \label{eq:app-order:rho-R}
    \rho_c(\omega,\vec{p})
    =
    \sum_{A=1}^{M+N-1}
    \sum_{\alpha=0}^{2(M+N)}
    \left( \frac{\partial}{\partial \omega_A} \right)^{\alpha}
    R_{A,\alpha}(\omega,\vec{p})
    \ ,
\end{gather}
with the definition
\begin{gather}
\label{eq:app-order:R}
R_{A,\alpha}(\omega,\vec{p})
=
\sum_{\beta=\alpha}^{2(M+N)}
\frac{[2(M+N)]!}{[2(M+N)-\beta]!\,[\beta-\alpha]!} m^\beta
\\ \nonumber \phantom{R_{A,\alpha}(\omega,\vec{p}) = } \times
\sum_{n_1,\dots,n_{M+N-1} \in \mathbb{Z}^4}
\mathring{G}_{n,A}(\omega,\vec{p}) \left( - \frac{\partial}{\partial \omega_A} \right)^{\beta-\alpha} w_n(\omega,\vec{p}) \ .
\end{gather}
Notice that the infinite sum over $n$ does not present issues of convergence since, when $(\omega,\vec{p})$ is restricted to any compact set, only a finite number of terms is different from zero. This representation shows explicitly that $R_{A,\alpha}(\omega,\vec{p})$ is locally $L^2$.

Eq.~\eqref{eq:app-order:rho-R} is the desired representation of the spectral density which satisfies eq.~\eqref{eq:app-order:N}. However, we are still left with the task of proving that $R_{A,\alpha}(\omega,\vec{p})$ is tempered. Since the map $p \mapsto (\omega,\vec{p})$ and its inverse are polynomially bounded and the Jacobian determinant is one, the temperedness of $R_{A,\alpha}(\omega,\vec{p})$ in the variables $(\omega,\vec{p})$ is equivalent to the temperedness of $R_{A,\alpha}(\omega,\vec{p})$ in the variables $p$. In order to show temperedness, it is therefore enough to prove that the following integral
\begin{gather}
    I_r = \int \Bigg[ \prod_{A=1}^{M+N-1} \frac{d^4p_A}{(2\pi)^4} \Bigg] \, \frac{\left| R_{A,\alpha}(\omega,\vec{p}) \right|}{ (3 + \| p \|_\infty )^r  }
\end{gather}
is finite for some value of $r \ge 0$. The derivative of $w_n(\omega,\vec{p})$ appearing in eq.~\eqref{eq:app-order:R} can be written in terms of $p$ using eq.~\eqref{eq:app-order:wn} and the linear relation between $\omega$ and $p_0$:
\begin{gather}
    \left( - \frac{\partial}{\partial \omega_A} \right)^{\beta-\alpha} w_n(\omega,\vec{p})
    =
    \left( \sum_{B=1}^{M+N-1} J_{AB} \frac{\partial}{\partial p_{B,0}} \right)^{\beta-\alpha}
    \prod_{B=1}^{M+N-1} \tilde{u}\left(\frac{p_B-n_B}{2}\right)
    \ ,
\end{gather}
for some constant matrix $J$. In particular, notice that the quantity
\begin{gather}
    K
    =
    \sum_{\beta=\alpha}^{2(M+N)} \!
    \frac{[2(M+N)]! \, m^\beta}{[2(M+N)-\beta]!\,[\beta-\alpha]!}
    \sup_{(\omega,\vec{p})} \left| \left( - \frac{\partial}{\partial \omega_A} \right)^{\beta-\alpha} w_n(\omega,\vec{p}) \right|
    \\ \nonumber =
    \sum_{\beta=\alpha}^{2(M+N)} \!
    \frac{[2(M+N)]! \, m^\beta}{[2(M+N)-\beta]!\,[\beta-\alpha]!}
    \sup_{p} \left| \left( \sum_{B=1}^{M+N-1} J_{AB} \frac{\partial}{\partial p_{B,0}} \right)^{\beta-\alpha}
    \prod_{B=1}^{M+N-1} \tilde{u}\left(\frac{p_B}{2}\right) \right|
\end{gather}
does not depend on $(n)$. Using this result, together with the observation that the closed support of $w_n(\omega,\vec{p})$ and all its derivatives is included in the set $\{ p \text{ s.t. } \| p-n \|_\infty \le 2 \}$, we obtain the following bound
\begin{gather}
    I_r
    \le
    K
    \sum_{n_1,\dots,n_{M+N-1} \in \mathbb{Z}^4}
    \int_{\| p-n \|_\infty \le 2} \Bigg[ \prod_{A=1}^{M+N-1} \frac{d^4p_A}{(2\pi)^4} \Bigg] \,
    \frac{ \left| \tilde{G}_{n,A}(q_0,\vec{p}) \right| }{ (3 + \| p \|_\infty )^r }
    \ ,
\end{gather}
where we have used eq.~\eqref{eq:app-order:Gring} to replace $\mathring{G}$ with $\tilde{G}$. The Cauchy-Schwartz inequality yields
\begin{gather}
    I_r
    \le
    K
    \sum_{n_1,\dots,n_{M+N-1} \in \mathbb{Z}^4}
    \left\{
    \int_{\| p-n \|_\infty \le 2} \Bigg[ \prod_{A=1}^{M+N-1} \frac{d^4p_A}{(2\pi)^4} \Bigg] \,
    \frac{ 1 }{ (3 + \| p \|_\infty )^{2r} }
    \right\}^{1/2}
    \\ \nonumber \hspace{35mm} \times
    \left\{
    \int_{\| p-n \|_\infty \le 2} \Bigg[ \prod_{A=1}^{M+N-1} \frac{d^4p_A}{(2\pi)^4} \Bigg] \,
    \left| \tilde{G}_{n,A}(q_0,\vec{p}) \right|^2
    \right\}^{1/2}
    \ .
\end{gather}
In the second integral, we drop the restriction on the integration domain and substitute the integration over $p_0$ with the integration over $q_0$. In the first integral, we use the bound
\begin{gather}
    3 + \| p \|_\infty \ge 3 + \| n \|_\infty - \| p-n \|_\infty \ge 1 + \| n \|_\infty \ ,
\end{gather}
which is valid in the integration domain. Therefore, we get
\begin{gather}
    I_r
    \le
    K
    \sum_{n_1,\dots,n_{M+N-1} \in \mathbb{Z}^4} \frac{2^{4(M+N-1)}}{ (1 + \| n \|_\infty )^r }
    \left\{
    \int \Bigg[ \prod_{A=1}^{M+N-1} \frac{dq_{A,0} d^3\vec{p}_A}{(2\pi)^4} \Bigg] \,
    \left| \tilde{G}_{n,A}(q_0,\vec{p}) \right|^2
    \right\}^{1/2}
    \nonumber \\
    \le
    K
    \sum_{n_1,\dots,n_{M+N-1} \in \mathbb{Z}^4} \frac{2^{4(M+N-1)}}{ (1 + \| n \|_\infty )^r }
    \left\{
    \int \Bigg[ \prod_{A=1}^{M+N-1} d\tau_A d^3\vec{x}_A \Bigg] \,
    \left| G_{n,A}(\tau,\vec{x}) \right|^2
    \right\}^{1/2}
    \ .
\end{gather}
In the last step we have used Parseval's identity. Using eqs.~\eqref{eq:app-order:FA}, \eqref{eq:app-order:bound-F-2} and \eqref{eq:app-order:G}, one obtains
\begin{gather}
    I_r
    \le
    K'
    \sum_{n_1,\dots,n_{M+N-1} \in \mathbb{Z}^4} \left| C_{n,\frac{3(M+N)}{2}} \right| \,(1 + \| n \|_\infty )^{-r}
    \ .
\end{gather}
where $K'$ is an $(n)$-independent finite constant given by:
\begin{gather}
    K' =
    K \, 2^{\frac{3(M+N)}{4}} 2^{4(M+N-1)}
    \left\{
    \int d^{M+N-1}\tau \,
    | 1 -i m \, \| \tau \|_\infty |^{-(M+N-1)-1}
    \right\}^{1/2}
    \\ \nonumber \hspace{35mm} \times
    \left\{
    \int d^{3(M+N-1)}\vec{x} \, ( 1 + m \| \vec{x} \|_\infty )^{-3(M+N-1)-1}
    \right\}^{1/2}
    \ .
\end{gather}
Momentarily, we will prove that $C_{n,\frac{3(M+N)}{2}}$ is bounded by a polynomial in $n$. Then, it is clear that $I_r$ can be made finite by choosing $r$ large enough. This proves the temperedness of $R_{A,\alpha}(\omega,\vec{p})$.

We use now the fact that $C_q(g)$ is a tempered distribution. Therefore, a continuous bounded function $\mathcal{C}(p)$, a polynomial $Q(p)$ and a multi-index $\gamma$ exist such that
\begin{gather}
    C_{\frac{3(M+N)}{2}}(g) = \int d^{4(M+N)}p \, Q(p) \mathcal{C}(p) D_p^\gamma \tilde{g}(p)
    \ .
\end{gather}
Using this representation with eq.~\eqref{eq:app-order:Cn}, we obtain
\begin{gather}
    C_{n,\frac{3(M+N)}{2}}
    \\ \nonumber =
    \int d^{4(M+N)}p \, Q(p) \mathcal{C}(p)
    \Bigg[ \prod_{A=1}^{M+N+1} \tilde{u}^{(\gamma_A)}(p_A-n_A) \Bigg] r_n^{-\|\gamma_{M+N}\|_1} \tilde{u}^{(\gamma_{M+N})} \left( \frac{p_{M+N}}{r_n} \right)
    \ .
\end{gather}
Replacing $p_A \to p_A+n_A$ for $A<M+N$ and $p_{M+N} \to r_n p_{M+N}$, and using the decomposition
\begin{gather}
    Q(p_1+n_1,\cdots,p_{M+N-1}+n_{M+N-1},r_n p_{M+N}) = \sum_{\ell=0}^L r_n^\ell Q'_\ell(n) Q''_\ell(p) \ ,
\end{gather}
where $Q'_\ell(n)$ and $Q''_\ell(p)$ are polynomials, we derive the following inequality
\begin{gather}
    \left| C_{n,\frac{3(M+N)}{2}} \right|
    \le
    \| \mathcal{C} \|_\infty \sum_{\ell=0}^L r_n^{4 -\|\gamma_{M+N}\|_1 + \ell} \, |Q'_\ell(n) |
    \int d^{4(M+N)}p \, \left| Q''_\ell(p) \, \prod_{A=1}^{M+N} \tilde{u}^{(\gamma_A)}(p_A) \right|
    \ .
\end{gather}
From here one easily proves that $\left| C_{n,\frac{3(M+N)}{2}} \right|$ is bounded by a polynomial in $n$, concluding the proof of the statement of this section.

\section{Results concerning the approximation}\label{app:approx}
In this appendix we use the same notation and definitions given in section~\ref{subsec:approx}. Our goal is to prove eq.~\eqref{eq:approx:bound}. We will do this through a number of technical steps.

\subsection{Finiteness of \texorpdfstring{$\bar{\omega}_A$}{ωbar}}
\label{app:approx:domain}

\textbf{Statement.} Define the set:
\begin{gather}
    \mathbb{E}_A = \bigg\{ \omega_A \text{ s.t. } q_{0,B} \ge E(\vec{q}_B) , \, \vec{p}_B \in \supp \check{f}_B \text{ for every } B
    \nonumber \\ \hspace{3cm}
    \text{and } {\textstyle \sum_{B=1}^{M+N} \eta_B \vec{p}_B } = 0
    \text{ and } {\textstyle \left| \sum_{B=1}^{M+N} \eta_B E(\vec{p}_B) \right| } \le \bar{\Delta}
    \bigg\}
    \label{eq:proofs-1:omegabar}
    \ .
\end{gather}
The constants $\bar{\omega}_A = \inf \mathbb{E}_A$ are well-defined and finite, i.e. $\mathbb{E}_A$ is non-empty and bounded from below.

\bigskip

\textbf{Proof.} Let us prove first that $\mathbb{E}_A$ is not empty. Recall that, in section~\ref{subsec:tantalo}, we required that some momenta $\bar{\vec{p}}_A$ exist which satisfy $\check{f}_A(\bar{\vec{p}}_A)\neq 0$ and the energy-momentum conservation conditions
\begin{gather}
    \sum_{A=1}^{M+N} \eta_A \bar{\vec{p}}_A = 0
    \ , \qquad
    \sum_{A=1}^{M+N} \eta_A E(\bar{\vec{p}}_A) = 0
    \ .
\end{gather}
It is easy to check that
\begin{gather}
    \label{eq:proofs-1:omega}
    \omega_A =
    \begin{cases}
        E\left( \sum_{B=1}^A \bar{\vec{p}}_B \right) - \sum_{B=1}^A E(\bar{\vec{p}}_B) \quad & \text{for } 1 \le A \le M \\[4pt]
        E\left( \sum_{B=M+1}^A \bar{\vec{p}}_B \right) - \sum_{B=M+1}^A E(\bar{\vec{p}}_B) \quad & \text{for } M<A<M+N
    \end{cases}
\end{gather}
is a non-zero element of $\mathbb{E}_A$. This implies that $\bar{\omega}_A$ is well defined and $\bar{\omega}_A < +\infty$.

Let us prove now that $\mathbb{E}_A$ is bounded from below. By loosening the conditions that define $\mathbb{E}_A$, one gets larger sets:
\begin{gather}
    \mathbb{E}_A \subseteq \mathbb{F}_A = \bigg\{ \omega_A \text{ s.t. } q_{0,B} \ge 0 , \, \vec{p}_B \in \supp \check{f}_B \text{ for every } B \bigg\}
    \ .
\end{gather}
Using eqs.~\eqref{eq:spectral:q} and \eqref{eq:spectral:omega}, one easily checks that if $\omega_A \in \mathbb{F}_A$ then
\begin{gather}
    \omega_A \ge - \sum_{B=1}^{M+N} E(\vec{p}_B) \ge - \sum_{B=1}^{M+N} \sup_{\vec{p}_B \in \supp \check{f}_B } E(\vec{p}_B) > -\infty
    \ ,
\end{gather}
where the finiteness follows from the fact that the support of $\check{f}_B$ is compact and $E(\vec{p}_B)$ is continuous.

\subsection{Existence of polynomial \texorpdfstring{$P_{\sigma,\epsilon}(z,\Delta)$}{P(z,Δ)}}
\label{app:approx:P}

\textbf{Statement.} Given $\epsilon>0$, we want to prove that a polynomial $P(z,\Delta)$ exists with the following properties: \textit{(1)} it vanishes if $z_A=0$ for any $A$, and $\textit{(2)}$ it satisfies the following bound:
\begin{gather}
    \max_{ \substack{ \| \alpha \|_1 \le \mathfrak{N}_\omega \\ 0 \le b \le \mathfrak{N}_{\vec{p}} } } \ \sup_{(\omega,\Delta) \in \mathbb{K}} \left| \mathcal{I}_{\alpha,b}(\omega,\Delta) \right| < \epsilon \ ,
    \label{eq:proofs-2:bound-uni}
\end{gather}
with the definition:
\begin{gather}
    \mathcal{I}_{\alpha,b}(\omega,\Delta) = \tau^{-\|\alpha\|_1} \bar{\Delta}^b
    \left[ \prod_{A=1}^{M+N-1} e^{\tau \omega_A} \partial_{\omega_A}^{\alpha_A} \right] \partial_\Delta^{b}
    \left[ K_\sigma(\omega,\Delta) - P(e^{-\tau \omega},\Delta) \right]
    \ .
    \label{eq:proofs-2:Ical}
\end{gather}
We recall that the integration domain is given by:
\begin{gather}
    \mathbb{K} = [\bar{\omega}_1{-}\bar{\Delta},+\infty) \times \cdots \times [\bar{\omega}_{M+N-1}{-}\bar{\Delta},+\infty) \times [ -\bar{\Delta},\bar{\Delta} ]
    \label{eq:proofs-2:Kbb}
    \ ,
\end{gather}
where the constants $\bar{\omega}_A$ are defined in eq.~\eqref{eq:approx:omegabar}. The polynomial $P(z,\Delta)$ also satisfies:
\begin{gather}
    \sum_{ \substack{ \| \alpha \|_1 \le \mathfrak{N}_\omega \\ 0 \le b \le \mathfrak{N}_{\vec{p}} } } \ \int_{\mathbb{K}} \frac{d^{M+N-1}\omega}{(2\pi)^{M+N-1}} d\Delta \, e^{-\tau \sum_{A=1}^{M+N-1} \omega_A} |\mathcal{I}_{\alpha,b}(\omega,\Delta)|^2
    < C \epsilon^2
    \ ,
    \label{eq:proofs-2:bound-L2}
\end{gather}
with the definition
\begin{gather}
C = 2\bar{\Delta} \frac{ e^{-\tau \sum_{A=1}^{M+N-1} (\bar{\omega}_A-\bar{\Delta})} }{(2\pi \tau)^{M+N-1} } (\mathfrak{N}_\omega+1)^{M+N-1} (\mathfrak{N}_{\vec{p}}+1) \ .
\end{gather}
Given the arbitrariness of $\epsilon$, eq.~\eqref{eq:proofs-2:bound-L2} implies eq.~\eqref{eq:approx:P}.

\bigskip

\textbf{Proof.} We define the auxiliary function
\begin{gather}
    X(z,\Delta) = \left[ \frac{K_\sigma(\omega,\Delta)}{z_1 \cdots z_{M+N-1}} \right]_{\omega_A = - \frac{1}{\tau} \log z_A} \ ,
\end{gather}
which is smooth with compact support in $(0,+\infty)^{N+M-1} \times \mathbb{R}$. Therefore, it can be extended by continuity to $[0,+\infty)^{N+M-1} \times \mathbb{R}$. Its extension, which will still be denoted by $X(z,\Delta)$, is a smooth function in $[0,+\infty)^{N+M-1} \times \mathbb{R}$ and vanishes with all its derivatives at the boundary of its domain.

We define the domain
\begin{gather}
    \mathbb{X} = [0,\bar{z}_1] \times \cdots \times [0,\bar{z}_{M+N-1}] \times [-\bar{\Delta},\bar{\Delta}]
    \label{eq:proofs-2:Xbb}
    \ , \\
    \bar{z}_A = e^{-\tau (\bar{\omega}_A-\bar{\Delta})}
    \label{eq:proofs-2:zbar-Deltabar}
    \ .
\end{gather}
We have already proved that $\bar{\omega}_A$ is finite (see appendix~\ref{app:approx:domain}), which implies $0 < \bar{z}_A < +\infty$. Therefore the set $\mathbb{X}$ is compact and it has a non-empty interior.

We start approximating the function $X(z,\Delta)$ in the compact domain $\mathbb{X}$, by means of the Bernstein polynomials (in shifted and rescaled variables)
\begin{gather}
    B_{n}(z,\Delta)
    =
    \sum_{k_1,\dots,k_{M+N-1}=0}^n \sum_{\ell=0}^n \left[ \prod_{A=1}^{M+N} \binom{n}{k_A} \left( \frac{z_A}{\bar{z}_A} \right)^{k_A} \left( 1-\frac{z_A}{\bar{z}_A} \right)^{n-k_A} \right]
    \\ \nonumber \phantom{B_{n}(z,\Delta) = } \times
    \binom{n}{\ell} \left( \frac{\bar{\Delta} + \Delta}{2\bar{\Delta}} \right)^{\ell} \left( \frac{\bar{\Delta} - \Delta}{2\bar{\Delta}} \right)^{n-\ell}
    X\left( \frac{\bar{z} k}{n} , -\bar{\Delta} +  \frac{2\bar{\Delta} \ell}{n} \right)
    \ ,
\end{gather}
which have the notable property that they approximate uniformly $X(z,\Delta)$ and all its derivatives (see e.g.~\cite{Martinez1989300, Foupouagnigni2020}), in particular:
\begin{gather}
    \lim_{n \to +\infty} \ \max_{ \substack{ \| \alpha \|_1 \le \mathfrak{N}_\omega \\ 0 \le b \le \mathfrak{N}_{\vec{p}} } } \ \sup_{(z,\Delta) \in \mathbb{X}} \ \bar{\Delta}^b \left| D_z^\alpha \partial_\Delta^b [ X(z,\Delta) - B_{n}(z,\Delta) ] \right| = 0
    \ .
\end{gather}
Therefore, for every $\delta>0$, a value of $n = \bar{n}(\delta)$ exists such that
\begin{gather}
    \max_{ \substack{ \| \alpha \|_1 \le \mathfrak{N}_\omega \\ 0 \le b \le \mathfrak{N}_{\vec{p}} } } \ \sup_{(z,\Delta) \in \mathbb{X}} \ \bar{\Delta}^b \left| D_z^\alpha \partial_\Delta^b [ X(z,\Delta) - B_{\bar{n}(\delta)}(z,\Delta) ] \right| < \delta
    \ .
    \label{eq:proofs-2:bound-1}
\end{gather}

We define the polynomial
\begin{gather}
    P(z,\Delta) = z_1 \cdots z_{M+N-1} B_{\bar{n}(\delta)}(z,\Delta) \ .
    \label{eq:proofs-2:P}
\end{gather}
Notice that $P(z,\Delta)$ vanishes if $z_A=0$ for any $A$. From now on, we assume the identification $z_A = e^{-\tau \omega_A}$.  We manipulate $\mathcal{I}_{\alpha,b}(\omega,\Delta)$ defined in eq.~\eqref{eq:proofs-2:Ical}. The chain and Leibniz rules imply the following identity
\begin{gather}
    e^{\tau \omega_A} \partial_{\omega_A}^{\alpha_A} g(z)
    =
    z_A^{-1} ( -\tau z_A \partial_{z_A})^{\alpha_A} g(z)
    \\ \nonumber
    =
    ( -\tau \partial_{z_A} z_A)^{\alpha_A} \frac{g(z)}{z_A}
    =
    ( -\tau)^{\alpha_A}
    \sum_{\alpha'_A=0}^{\alpha_A} c_{\alpha'_A}^{\alpha_A} z_A^{\alpha'_A} \partial_{z_A}^{\alpha'_A} \frac{g(z)}{z_A}
    \ ,
\end{gather}
for some constants $c_{\alpha'_A}^{\alpha_A}$ which can be calculated recursively. When using this expression in eq.~\eqref{eq:proofs-2:Ical}, with the definition of the functions $X(z,\Delta)$ and $P(z,\Delta)$ one readily obtains:
\begin{gather}
    \mathcal{I}_{\alpha,b}(\omega,\Delta) = (-1)^{\|\alpha\|_1} \bar{\Delta}^b
    \sum_{\alpha' \le \alpha}
    \left[ \prod_{A=1}^{M+N-1}
    c_{\alpha'_A}^{\alpha_A} z_A^{\alpha'_A} \partial_{z_A}^{\alpha'_A}
    \right] \partial_\Delta^{b}
    \left[ X(z,\Delta) - B_{\bar{n}(\delta)}(z,\Delta) \right]
    \ .
\end{gather}
Since the image of $\mathbb{K}$ under the change of variables $(\omega,\Delta) \mapsto (z,\Delta)$ is precisely $\mathbb{X}$, eq.~\eqref{eq:proofs-2:bound-1} yields
\begin{gather}
    \label{eq:proofs-2:bound-2}
    \max_{ \substack{ \| \alpha \|_1 \le \mathfrak{N}_\omega \\ 0 \le b \le \mathfrak{N}_{\vec{p}} } } \ \sup_{(\omega,\Delta) \in \mathbb{K}}
    \left| \mathcal{I}_{\alpha,b}(\omega,\Delta) \right|
    =
    \max_{ \substack{ \| \alpha \|_1 \le \mathfrak{N}_\omega \\ 0 \le b \le \mathfrak{N}_{\vec{p}} } } \ \sup_{(z,\Delta) \in \mathbb{X}}
    \left| \mathcal{I}_{\alpha,b}(\omega,\Delta) \right|
    \\ \nonumber <
    \delta \,
    \max_{ \| \alpha \|_1 \le \mathfrak{N}_\omega }
    \sum_{\alpha' \le \alpha}
    \prod_{A=1}^{M+N-1} |c_{\alpha'_A}^{\alpha_A}| \bar{z}_A^{\alpha'_A}
    \ .
\end{gather}
Since $\delta$ is arbitrary, we can choose
\begin{gather}
    \delta = \left[ \max_{ \| \alpha \|_1 \le \mathfrak{N}_\omega }
    \sum_{\alpha' \le \alpha}
    \prod_{A=1}^{M+N-1} |c_{\alpha'_A}^{\alpha_A}| \bar{z}_A^{\alpha'_A} \right]^{-1} \epsilon
    \ .
\end{gather}
With this choice, eq.~\eqref{eq:proofs-2:bound-2} is nothing but eq.~\eqref{eq:proofs-2:bound-uni}. Finally notice that
\begin{gather}
    \sum_{ \substack{ \| \alpha \|_1 \le \mathfrak{N}_\omega \\ 0 \le b \le \mathfrak{N}_{\vec{p}} } } \ \int_{\mathbb{K}} \frac{d^{M+N-1}\omega}{(2\pi)^{M+N-1}} d\Delta \, e^{-\tau \sum_{A=1}^{M+N-1} \omega_A} |\mathcal{I}_{\alpha,b}(\omega,\Delta)|^2
    \\ \nonumber <
    \epsilon^2 \sum_{ \substack{ \| \alpha \|_1 \le \mathfrak{N}_\omega \\ 0 \le b \le \mathfrak{N}_{\vec{p}} } } \ \int_{\mathbb{K}} \frac{d^{M+N-1}\omega}{(2\pi)^{M+N-1}} d\Delta \, e^{-\tau \sum_{A=1}^{M+N-1} \omega_A}
    \\ \nonumber \le
    2\bar{\Delta} \frac{ e^{-\tau \sum_{A=1}^{M+N-1} (\bar{\omega}_A-\bar{\Delta})} }{(2\pi \tau)^{M+N-1} } (\mathfrak{N}_\omega+1)^{M+N-1} (\mathfrak{N}_{\vec{p}}+1) \epsilon^2
    \ .
\end{gather}
which is eq.~\eqref{eq:proofs-2:bound-L2}.

\subsection{Error of approximation}
\label{app:approx:error}

\textbf{Statement.} For every $\sigma>0$ and $\epsilon > 0$, the approximation $\mathcal{S}_c(\sigma,\epsilon)$ constructed in subsection~\ref{subsec:approx} satisfies the following bound:
\begin{gather}
    \left| \mathcal{S}_c(\sigma,\epsilon) - \mathcal{S}_c(\sigma) \right| < a \epsilon \ ,
    \label{eq:proofs:bound}
\end{gather}
for some constant $a$ which depends on all data of the problem but $\epsilon$ and $\sigma$. By combining this fact with eq.~\eqref{eq:tantalo:TS}, the bound in eq.~\eqref{eq:approx:bound} follows.

\bigskip

\textbf{Proof.} First, we rewrite the connected transition amplitude given in eq.~\eqref{eq:spectral:S-sigma} more compactly as:
\begin{gather}
    \mathcal{S}_c(\sigma)
    =
    \int d\mu(\omega,\vec{p}) \, \Xi(\vec{p}) \, K_\sigma(\omega,\mathring{\Delta}(\vec{p})) \, \rho_c(\omega,\vec{p})
    \ .
    \label{eq:proofs:S-sigma}
\end{gather}
In going from eq.~\eqref{eq:spectral:S-sigma} to the above representation we have used the delta of momentum conservation to remove the integral over $\vec{p}_{M+N}$ and we have defined
\begin{gather}
    \Xi(\vec{p})
    =
    \Bigg[ \prod_{A=1}^{M+N-1} \check{f}_A^{(*)}(\vec{p}_A) \Bigg]
    \check{f}_{M+N}^*\left( - \sum_{A=1}^{M+N-1} \eta_A \vec{p}_A \right)
    \, \tilde{h}(\mathring{\Delta}(\vec{p}))
    \ , \\
    \mathring{\Delta}(\vec{p}) = \sum_{B=1}^{M+N-1} \eta_B E(\vec{p}_B) + E\left( - \sum_{A=1}^{M+N-1} \eta_A \vec{p}_A \right)
    \ , \\
    d\mu(\omega,\vec{p}) = \prod_{A=1}^{M+N-1} \frac{d \omega_A d^3 \vec{p}_A}{(2\pi)^4}
    \ .
\end{gather}
The transition amplitude $\mathcal{S}_c(\sigma,\epsilon)$ is obtained by replacing $K_\sigma(\omega,\Delta)$ in eq.~\eqref{eq:proofs:S-sigma} with its approximation $P_{\sigma,\epsilon}(e^{-\tau \omega},\Delta)$ constructed in subsection~\ref{subsec:approx}, i.e.
\begin{gather}
    \mathcal{S}_c(\sigma,\epsilon) =
    \int d\mu(\omega,\vec{p}) \, \Xi(\vec{p}) \, P_{\sigma,\epsilon}(e^{-\tau \omega},\mathring{\Delta}(\vec{p})) \, \rho_c(\omega,\vec{p})
    \ .
    \label{eq:proofs:S-sigma-eps}
\end{gather}

Eqs.~\eqref{eq:spectral:suppWc} and \eqref{eq:spectral:rho} imply that the support of the $\Xi(\vec{p}) \rho_c(\omega,\vec{p})$ is a subset of $\mathbb{D}_\omega \times \mathbb{D}_{\vec{p}}$ defined by:
\begin{gather}
    \mathbb{D}_\omega = [\bar{\omega}_1,+\infty) \times \cdots \times [\bar{\omega}_{M+N-1},+\infty)
    \label{eq:proofs:Dbb-omega}
    \ , \\
    \mathbb{D}_{\vec{p}} = \bigg\{ (\vec{p}_1,\dots,\vec{p}_{M+N-1}) \text{ s.t. } \vec{p}_B \in \supp \check{f}_B \text{ for every } B=1,\dots,M+N
    \nonumber \\ \hspace{3cm}
    \text{and } {\textstyle \left| \sum_{B=1}^{M+N} \eta_B E(\vec{p}_B) \right| } \le \bar{\Delta}
    \text{ and } {\textstyle \sum_{B=1}^{M+N} \eta_B \vec{p}_B } = 0
    \bigg\}
    \label{eq:proofs:Dbb-p}
    \ , \\
    \bar{\omega}_A = \inf \bigg\{ \omega_A \text{ s.t. } q_{0,B} \ge E(\vec{q}_B) \text{ for every } B \text{ and } \vec{p} \in \mathbb{D}_{\vec{p}} \bigg\}
    \ .
\end{gather}
Notice that this definition of $\bar{\omega}_A$ is equivalent to eq.~\eqref{eq:approx:omegabar}. We define also the set
\begin{gather}
    \mathbb{D}'_\omega = [\bar{\omega}_1{-}\bar{\Delta},+\infty) \times \cdots \times [\bar{\omega}_{M+N-1}{-}\bar{\Delta},+\infty)
    \label{eq:proofs:Dbbprime-omega}
\end{gather}
and we observe that $\mathbb{K} = \mathbb{D}'_\omega \times [-\bar{\Delta},\bar{\Delta}]$, where $\mathbb{K}$ is defined in eq.~\eqref{eq:approx:Kbb}. We introduce an auxiliary smooth function $\vartheta_A(\omega)$ with the following properties: \textit{(1)} $0 \le \vartheta_A(\omega) \le 1$, \textit{(2)} $\vartheta(\omega)=1$ for $\omega \ge \bar{\omega}_A - \bar{\Delta}/2$, and \textit{(3)} $\vartheta(\omega)=0$ for $\omega \le \bar{\omega}_A-\bar{\Delta}$. We define
\begin{gather}
    \Xi'(\omega,\vec{p}) = \Xi(\vec{p}) \prod_{A=1}^{M+N-1} \vartheta(\omega_A) \ .
\end{gather}
By construction, this function satisfies
\begin{gather}
    \Xi'(\omega,\vec{p}) \rho_c(\omega,\vec{p}) = \Xi(\vec{p}) \rho_c(\omega,\vec{p})
    \ , \qquad
    \supp \Xi' \subseteq \mathbb{D}'_\omega \times \mathbb{D}_{\vec{p}}
    \ .
\end{gather}
Since $\Xi' \rho_c = \Xi \rho_c$, we can freely replace $\Xi(\vec{p})$ with $\Xi'(\omega,\vec{p})$ in eqs.~\eqref{eq:proofs:S-sigma} and \eqref{eq:proofs:S-sigma-eps}. Then, the representation of $\rho_c(q)$ as the sum of weak derivatives of some $L^2$-tempered functions given by eq.~\eqref{eq:order:rho-R} yields the bound:
\begin{gather}
    \label{eq:proofs:error-1}
    | \mathcal{S}_c(\sigma,\epsilon) - \mathcal{S}_c(\sigma) |
    \\ \nonumber =
    \left| \int d\mu(\omega,\vec{p}) \, \Xi'(\omega,\vec{p}) \,  Z(\omega,\mathring{\Delta}(\vec{p})) \, \rho_c(\omega,\vec{p}) \right|
    \\ \nonumber \le
    \sum_{ \substack{\vphantom{\beta}\alpha \text{ s.t.} \\ \|\alpha\|_1 \le \mathfrak{N}_\omega} }
    \sum_{ \substack{\beta \text{ s.t.} \\ \|\beta\|_1 \le \mathfrak{N}_{\vec{p}}} }
    \int d\mu(\omega,\vec{p}) \, | R_{\alpha,\beta}(\omega,\vec{p}) | \, \left| D_\omega^\alpha D_{\vec{p}}^\beta \left[ \Xi'(\omega,\vec{p}) Z(\omega,\mathring{\Delta}(\vec{p})) \right] \right|
    \ ,
\end{gather}
with the definition
\begin{gather}
    Z(\omega,\Delta) = K_\sigma(\omega,\Delta) - P_{\sigma,\epsilon}(e^{-\tau \omega},\Delta)
    \ .
\end{gather}

We focus on the integrand in the last expression of eq.~\eqref{eq:proofs:error-1}. The Leibniz rule allows us to write
\begin{gather}
    \label{eq:proofs:DXW}
    D_\omega^\alpha D_{\vec{p}}^\beta \left[ \Xi'(\omega,\vec{p}) Z(\omega,\mathring{\Delta}(\vec{p})) \right]
    \\ \nonumber \le
    \sum_{\substack{\vphantom{\beta}\alpha' \text{ s.t.} \\ \vphantom{\beta}\alpha'_A \le \alpha_A}}
    \sum_{\substack{\beta' \text{ s.t.} \\ \beta'_{A,k} \le \beta_{A,k}}}
    c_{\alpha',\beta'}^{\alpha,\beta}
    D_\omega^{\alpha-\alpha'} D_{\vec{p}}^{\beta-\beta'} \Xi'(\omega,\vec{p})
    D_\omega^{\alpha'} D_{\vec{p}}^{\beta'} Z(\omega,\mathring{\Delta}(\vec{p}))
    \ ,
\end{gather}
for certain coefficients $c_{\alpha',\beta'}^{\alpha,\beta}$. In the above formula, $\alpha_A$ and $\alpha'_A$ are multi-indices with $A=1,\dots,M+N-1$, $\beta_{A,k}$ and $\beta'_{A,k}$ are multi-indices with $A=1,\dots,M+N-1$ and $k=1,2,3$. The chain rule implies the following formula
\begin{gather}
    D_{\vec{p}}^{\beta'} Z(\omega,\mathring{\Delta}(\vec{p}))
    =
    \sum_{\gamma'=0}^{\|\beta'\|_1}
    u^{\beta'}_{\gamma'}(\vec{p}) \partial_\Delta^{\gamma'} Z(\omega,\mathring{\Delta}(\vec{p}))
    \ ,
    \label{eq:proofs:DX}
\end{gather}
for some smooth functions $u^{\beta'}_{\gamma'}(\vec{p})$ which are polynomially-bounded with all their derivatives, and depend on no other data of the problem except $M$, $N$ and $\tau$. The above formula can be proven by induction on $\beta'$. The formula clearly works for $\beta'=(0,\dots,0)$ with:
\begin{gather}
    u^{(0,\dots,0)}_{\gamma'}(\vec{p}) = \delta_{\gamma',0} \ .
\end{gather}
By differentiating both sides of the formula and using standard properties of derivatives, one obtains the following recursive relation:
\begin{gather}
    u^{\beta'+e^{(A,k)}}_{\gamma'}(\vec{p})
    =
    \frac{\partial u^{\beta'}_{\gamma'}}{\partial p_{A,k}}(\vec{p})
    + u^{\beta'}_{\gamma'-1}(\vec{p}) \frac{\partial \mathring{\Delta}}{\partial p_{A,k}}(\vec{p})
    \ ,
\end{gather}
where we have defined the multi-indices $e^{(A,k)}_{B,\ell} = \delta_{AB} \delta_{k\ell}$. Notice that $\mathring{\Delta}(\vec{p})$ is smooth and polynomially-bounded with all its derivatives. The corresponding property of $u^{\beta'}_{\gamma}(\vec{p})$ follows by induction.
Using eqs.~\eqref{eq:proofs:DXW} and \eqref{eq:proofs:DX} with eq.~\eqref{eq:proofs:error-1}, and rearranging judiciously the sums, one obtains the bound
\begin{gather}
    \label{eq:proofs:error-2}
    | \mathcal{S}_c(\sigma,\epsilon) - \mathcal{S}_c(\sigma) |
    \\ \nonumber \le
    \sum_{\gamma'=0}^{\mathfrak{N}_{\vec{p}}}
    \sum_{ \substack{\alpha' \text{ s.t.} \\ \|\alpha'\|_1 \le \mathfrak{N}_\omega} }
    \int_{\mathbb{D}'_\omega \times \mathbb{D}_{\vec{p}}} d\mu(\omega,\vec{p}) \,
    R'_{\alpha',\gamma'}(\omega,\vec{p}) \,
    \left| D_\omega^{\alpha'} \partial_\Delta^{\gamma'} Z(\omega,\mathring{\Delta}(\vec{p})) \right|
    \ ,
\end{gather}
with the definition
\begin{gather}
    R'_{\alpha',\gamma'}(\omega,\vec{p})
    = \!\!
    \sum_{\substack{\vphantom{\beta'}\alpha \text{ s.t.} \\ \vphantom{\beta'_{A,k}}\alpha'_A \le \alpha_A  \\ \vphantom{\beta'} \|\alpha\|_1 \le \mathfrak{N}_\omega}} \,
    \sum_{\substack{\beta, \beta' \text{ s.t.} \\ \beta'_{A,k} \le \beta_{A,k} \\ \gamma' \le \|\beta'\|_1 \le \| \beta\|_1 \le \mathfrak{N}_{\vec{p}}}} \!\!\!\!\!\!\!\!\!
    \left|
    c_{\alpha',\beta'}^{\alpha,\beta}
    u^{\beta'}_{\gamma'}(\vec{p})
    R_{\alpha,\beta}(\omega,\vec{p})
    D_\omega^{\alpha-\alpha'} D_{\vec{p}}^{\beta-\beta'} \Xi'(\omega,\vec{p})
    \right|
    \ .
    \label{eq:proofs:R-1}
\end{gather}
The restriction of the integral in eq.~\eqref{eq:proofs:error-2} is allowed because all derivatives of $\Xi'(\omega,\vec{p})$ and, hence, the function $R'_{\alpha',\gamma'}(\omega,\vec{p})$ have support in $\mathbb{D}'_\omega \times \mathbb{D}_{\vec{p}}$. Notice that the function $R'_{\alpha',\gamma'}(\omega,\vec{p})$ is $L^2$-tempered.

We introduce the integral operators $I_A$ :
\begin{gather}
    I_A g(\dots,\omega_A,\dots) = \int_{\omega_A}^{+\infty} dy \, g(\dots,y,\dots)
    \ , \\
    J_A g(\dots,\omega_A,\dots) = \int_{\bar{\omega}_A-\bar{\Delta}}^{\omega_A} dy \, g(\dots,y,\dots)
    \ ,
\end{gather}
and we notice that
\begin{gather}
    \label{eq:proofs:IJ}
    \int_{\mathbb{K}} d\mu(\omega,\vec{p}) \, g_1(\omega,\mathring{\Delta}(\vec{p})) I_A g_2(\omega,\mathring{\Delta}(\vec{p}))
    =
    \int_{\mathbb{K}} d\mu(\omega,\vec{p}) \, J_A g_1(\omega,\mathring{\Delta}(\vec{p})) g_2(\omega,\mathring{\Delta}(\vec{p}))
    \ ,
\end{gather}
as long as the function $g_1(\omega,\Delta) g_2(\omega_1,\dots,\omega_{A-1},y,\omega_{A+1},\dots,\omega_{M+N-1},\Delta) \theta(y-\omega_A)$ is integrable in all its variables. For any multi-index $\alpha'$ satisfying $\|\alpha'\|_1 \le \mathfrak{N}_\omega$, one can find a multi-index $\bar{\alpha}'$ with the following two properties: \textit{(1)} $\alpha'_A \le \bar{\alpha}'_A$ for every $A$, and \textit{(2)} $\| \bar{\alpha}' \|_1 = \mathfrak{N}_\omega$. Then, the following identity
\begin{gather}
    D_\omega^{\alpha'} \partial_\Delta^{\gamma'} Z(\omega,\mathring{\Delta}(\vec{p}))
    =
    (-1)^{\|\bar{\alpha}'-\alpha'\|_1} I^{\bar{\alpha}'-\alpha'} D_\omega^{\bar{\alpha}'} \partial_\Delta^{\gamma'} Z(\omega,\mathring{\Delta}(\vec{p}))
\end{gather}
follows from a recursive use of integration by parts with the observation that $Z(\omega,\mathring{\Delta}(\vec{p}))$ and all its derivatives vanish in the $\omega_A \to +\infty$ limit for any $A$. Then, the trivial inequality $|I_A g| \le I_A |g|$ yields
\begin{gather}
    \left| D_\omega^{\alpha'} \partial_\Delta^{\gamma'} Z(\omega,\mathring{\Delta}(\vec{p})) \right|
    \le
    I^{\bar{\alpha}'-\alpha'} \left| D_\omega^{\bar{\alpha}'} \partial_\Delta^{\gamma'} Z(\omega,\mathring{\Delta}(\vec{p})) \right|
\end{gather}
We plug this back into eq.~\eqref{eq:proofs:error-2} and use eq.~\eqref{eq:proofs:IJ}:
\begin{gather}
    \label{eq:proofs:error-3}
    | \mathcal{S}_c(\sigma,\epsilon) - \mathcal{S}_c(\sigma) |
    \le
    \sum_{\gamma'=0}^{\mathfrak{N}_{\vec{p}}}
    \sum_{ \substack{\alpha'' \text{ s.t.} \\ \|\alpha''\|_1 = \mathfrak{N}_\omega} }
    \int_{\mathbb{D}'_\omega \times \mathbb{D}_{\vec{p}}} \!\!\! d\mu(\omega,\vec{p}) \,
    R''_{\alpha'',\gamma'}(\omega,\vec{p}) \,
    \left| D_\omega^{\alpha''} \partial_\Delta^{\gamma'} Z(\omega,\mathring{\Delta}(\vec{p})) \right|
    \ ,
\end{gather}
with the definition
\begin{gather}
    R''_{\alpha'',\gamma'}(\omega,\vec{p})
    =
    \sum_{ \substack{\alpha' \text{ s.t.} \\ \|\alpha'\|_1 \le \mathfrak{N}_\omega \\ \bar{\alpha}' = \alpha''} }
    J^{\bar{\alpha}'-\alpha'} R'_{\alpha',\gamma'}(\omega,\vec{p})
    \ .
    \label{eq:proofs:Rprime}
\end{gather}
It is easy to show that $J_A$ maps the set of $L^2$-tempered functions into itself. Therefore, like $R'_{\alpha',\gamma'}(\omega,\vec{p})$, the function $R''_{\alpha',\gamma'}(\omega,\vec{p})$ is $L^2$-tempered. Applying the Cauchy-Schwartz inequality to eq.~\eqref{eq:proofs:error-3}, we obtain:
\begin{gather}
    \label{eq:proofs:error-4}
    | \mathcal{S}_c(\sigma,\epsilon) - \mathcal{S}_c(\sigma) |
    \\ \nonumber \le
    a' \left\{
    \sum_{\gamma'=0}^{\mathfrak{N}_{\vec{p}}}
    \sum_{ \substack{\alpha'' \text{ s.t.} \\ \|\alpha''\|_1 = \mathfrak{N}_\omega} }
    \bar{\Delta}^{\gamma'}
    \int_{\mathbb{D}'_\omega \times \mathbb{D}_{\vec{p}}} d\mu(\omega,\vec{p}) \,
    e^{\tau \sum_{A=1}^{M+N-1} \omega_A}
    \left| D_\omega^{\alpha''} \partial_\Delta^{\gamma'} Z(\omega,\mathring{\Delta}(\vec{p})) \right|^2 \right\}^{1/2}
    \ ,
\end{gather}
where the constant
\begin{gather}
    \label{eq:proofs:aprime}
    a' =
    \left\{
    \sum_{\gamma'=0}^{\mathfrak{N}_{\vec{p}}}
    \sum_{ \substack{\alpha'' \text{ s.t.} \\ \|\alpha''\|_1 = \mathfrak{N}_\omega} }
    \bar{\Delta}^{-\gamma'}
    \int_{\mathbb{D}'_\omega \times \mathbb{D}_{\vec{p}}} d\mu(\omega,\vec{p}) \,
    e^{-\tau \sum_{A=1}^{M+N-1} \omega_A}
    | R''_{\alpha'',\gamma'}(\omega,\vec{p}) |^2
    \right\}^{1/2}
\end{gather}
is finite because $\mathbb{D}_{\vec{p}}$ is compact and $R''_{\alpha',\gamma'}(\omega,\vec{p})$ is an $L^2$-tempered function. We want to find some inequality which allows us to replace the integration over $\vec{p}_A$ in eq.~\eqref{eq:proofs:error-4} with an integration over $\Delta = \mathring{\Delta}(\vec{p})$. We will do this in a few steps.

We are interested the integral
\begin{gather}
    \label{eq:cov:I-1}
    \mathcal{I} = \int_{\mathbb{D}_{\vec{p}}} \Bigg[ \prod_{A=1}^{M+N-1} \frac{d^3 \vec{p}_A}{(2\pi)^3} \Bigg] \, F(\mathring{\Delta}(\vec{p}))
\end{gather}
where $F(\Delta) = \left| D_\omega^{\alpha''} \partial_\Delta^{\gamma'} Z(\omega,\Delta) \right|^2$ is a smooth function. The $\omega$ dependence does not play any role here and we suppress it for brevity. A few lines of algebra show that the above integral can be written equivalently as
\begin{gather}
    \label{eq:cov:I-2}
    \mathcal{I} = \int dP^0_\sin \, dP^0_\sout \, d^3\vec{P} \,
    w_\sin(P^0_\sin,\vec{P}) \, w_\sout(P^0_\sout,\vec{P}) \,
    F( P^0_\sout - P^0_\sin ) \, \chi_{[-\bar{\Delta},\bar{\Delta}]}( P^0_\sout - P^0_\sin )
    \ ,
\end{gather}
with the definitions
\begin{gather}
    \label{eq:cov:win-1}
    w_\sin(P) = \int_{\vec{p}_B \in \supp \check{f}_B} \Bigg[ \prod_{A=1}^{M} \frac{d^3\vec{p}_A}{(2\pi)^3} \Bigg] \, \delta^4\Bigg( P - \sum_{A=1}^{M} p_A \Bigg)_{p_A^0 = E(\vec{p}_A)}
    \ , \\
    \label{eq:cov:wout-1}
    w_\sout(P) = \int_{\vec{p}_B \in \supp \check{f}_B} \Bigg[ \prod_{A=M+1}^{M+N} \frac{d^3\vec{p}_A}{(2\pi)^3} \Bigg] \, \delta^4\Bigg( P - \sum_{A=M+1}^{M+N} p_A \Bigg)_{p_A^0 = E(\vec{p}_A)}
    \ .
\end{gather}
From this representation it is clear that $w_\sin(P)$ and $w_\sout(P)$ are non-negative distributions, whose support satisfies
\begin{gather}
    \supp w_\as \subset \mathbb{W} = \{ P \in \mathbb{R}^4 \text{ s.t. } 0 < P^0 < \mathcal{E} \text{ and } P^2>0 \} \ ,
\end{gather}
for $\as = \sin, \sout$, with the definitions
\begin{gather}
    \mathcal{E} = \bar{\Delta} + \sup_{\vec{p}_A \in \supp \check{f}_A} \max \left\{ \sum_{A=1}^M E(\vec{p}_A) , \sum_{A=M+1}^{M+N} E(\vec{p}_A) \right\}
    \ ,
\end{gather}
which is finite and larger than zero, because all wave functions have non-empty compact support. We will prove that a positive constant $C$ exists such that
\begin{gather}
    \label{eq:cov:wbounds}
    w_\as(P) \le C \chi_{\mathbb{W}}(P) \ .
\end{gather}
In particular, this bound implies that $w_\sin(P)$ and $w_\sout(P)$ are functions rather than distributions,\footnote{
Let $T(x)$ be a tempered distribution satisfying $0 \le T(x) \le a$ for some $a > 0$. Since $T(x)$ is positive, it is a tempered Radon measure (see exercise 4 in chapter 6 in~\cite{rudin1974functional}). Hence, it can be applied to any continuous function with compact support. Let $f(x)$ one of such functions. Let $f_+(x)$ and $f_-(x)$ be the positive and negative part of $f(x)$. Clearly, $f_+(x)$ and $f_-(x)$ are both continuous functions with compact support. Therefore
\begin{gather}
    |(T,f)| \le |(T,f_+)| + |(T,f_-)| =  (T,f_+) - (T,f_-) = (T,|f|)
    \le a \|f\|_1 \ .
\end{gather}
Therefore $T$ is in the dual of $L^1$, i.e. it is an element of $L^\infty$ and, in particular, a measurable function.
}%
which makes the integral in eq.~\eqref{eq:cov:I-2} well defined. By using eq.~\eqref{eq:cov:wbounds} in eq.~\eqref{eq:cov:I-2} and enlarging the integration domain, we obtain
\begin{gather}
    \label{eq:cov:I-3}
    \mathcal{I} \le C^2 \int_0^{\mathcal{E}} dP^0_\sin \int_0^{\mathcal{E}} dP^0_\sout \int_{\vec{P}^2 \le \mathcal{E}^2} d^3\vec{P}
    F( P^0_\sout - P^0_\sin ) \, \chi_{[-\bar{\Delta},\bar{\Delta}]}( P^0_\sout - P^0_\sin )
    \\ \nonumber \phantom{\mathcal{I}} \le
    \frac{4 \pi C^2 \mathcal{E}^3}{3} \int_0^{\mathcal{E}} dP^0_\sin \int_0^{\mathcal{E}} dP^0_\sout
    F( P^0_\sout - P^0_\sin ) \, \chi_{[-\bar{\Delta},\bar{\Delta}]}( P^0_\sout - P^0_\sin )
    \ .
\end{gather}
We change variables to $P^0 = (P^0_\sin + P^0_\sout)/2$ and $\Delta = P^0_\sout - P^0_\sin$ and we enlarge the integration domain once more, obtaining
\begin{gather}
    \label{eq:cov:I-4}
    \mathcal{I} \le \frac{4 \pi C^2 \mathcal{E}^3}{3}
    \int_0^{\mathcal{E}} dP^0 \int_{-\bar{\Delta}}^{\bar{\Delta}} d\Delta \, F( \Delta )
    =
    \frac{4 \pi C^2 \mathcal{E}^4}{3} \int_{-\bar{\Delta}}^{\bar{\Delta}} d\Delta \, F( \Delta )
    \ .
\end{gather}
Using this inequality in eq.~\eqref{eq:proofs:error-4} and observing that $\mathbb{D}'_\omega \times [-\bar{\Delta},\Delta] = \mathbb{K}$ defined in eq.~\eqref{eq:approx:Kbb}, we obtain:
\begin{gather}
    \label{eq:proofs:error-5}
    | \mathcal{S}_c(\sigma,\epsilon) - \mathcal{S}_c(\sigma) |
    \\ \nonumber \le
    a \left\{
    \sum_{\gamma'=0}^{\mathfrak{N}_{\vec{p}}}
    \sum_{ \substack{\alpha'' \text{ s.t.} \\ \|\alpha''\|_1 = \mathfrak{N}_\omega} }
    \bar{\Delta}^{\gamma'}
    \int_{\mathbb{K}} \Bigg[ \prod_{A=1}^{M+N-1} \frac{d \omega_A}{2\pi} \Bigg] d\Delta \,
    e^{\tau \sum_{A=1}^{M+N-1} \omega_A}
    \left| D_\omega^{\alpha''} \partial_\Delta^{\gamma'} Z(\omega,\Delta) \right|^2 \right\}^{1/2}
    \ ,
\end{gather}
with the definition
\begin{gather}
    a =
    \frac{4 \pi C^2 \mathcal{E}^4}{3} a' \ .
\end{gather}
One sees explicitly that $a$ is independent of $\epsilon$ and $\sigma$. Finally, eq.~\eqref{eq:approx:P} with eq.~\eqref{eq:proofs:error-5} yields
\begin{gather}
    \label{eq:proofs:error-6}
    | \mathcal{S}_c(\sigma,\epsilon) - \mathcal{S}_c(\sigma) | \le a \epsilon \ .
\end{gather}

The asymptotic behavior given in eq.~\eqref{eq:tantalo:TS} together with the observation that $\mathcal{S}_c(\sigma)$ is bounded, implies that a constant $b_r$ (which is independent of $\sigma$ and, clearly, $\epsilon$) exists such that, for any $\sigma>0$
\begin{gather}
    \left| \mathcal{S}_c(\sigma) - S_c \right| < b_r \sigma^r \ ,
\end{gather}
and, therefore,
\begin{gather}
    \left| \mathcal{S}_c(\sigma,\epsilon) - S_c \right| \le \left| \mathcal{S}_c(\sigma,\epsilon) - \mathcal{S}_c(\sigma) \right| + \left| \mathcal{S}_c(\sigma) - S_c \right| <  a \epsilon + b_r \sigma^r \ ,
\end{gather}
which is exactly what we wanted to prove.

We are left with the task to prove eq.~\eqref{eq:cov:wbounds}. Let us focus on $w_\sin(P)$. The (closed) support of $w_\sin(P)$ is a subset of the open set
\begin{gather}
    \mathbb{W}_\sin = \{ P \in \mathbb{R}^4 \text{ s.t. } 0 < P^0 < 2\mathcal{E}_\sin \text{ and } P^2>0 \} \ .
\end{gather}
In order to prove that $w_\sin(P)$ is bounded from above, it is enough to prove boundedness for $P \in \mathbb{W}_\sin$. In this case, we use the inequality
\begin{gather}
    1 = \prod_{A=1}^M \frac{2E(\vec{p}_A)}{2E(\vec{p}_A)} \le
    (4\mathcal{E}_\sin)^M
    \prod_{A=1}^M \frac{1}{2E(\vec{p}_A)}
    \ ,
\end{gather}
which yields
\begin{gather}
    \label{eq:cov:win-2}
    w_\sin(P) \le (4\mathcal{E}_\sin)^M \int \Bigg[ \prod_{A=1}^{M} \frac{d^3\vec{p}_A}{(2\pi)^3 2E(\vec{p}_A)} \Bigg] \, \delta^4\left( P - \sum_{A=1}^{M} p_A \right)_{p_A^0 = E(\vec{p}_A)}
    \ .
\end{gather}
In deriving this inequality, we have also enlarged the integration domain over $\vec{p}_A$ from $\supp \check{f}_A$ to the whole space. The right-hand side of the above inequality is invariant under Lorentz transformations of $P$. As long as $P$ is in $\mathbb{W}_\sin$, the integral can be calculated in the frame defined by $\vec{P}=\vec{0}$, i.e.
\begin{gather}
    \label{eq:cov:win-3}
    w_\sin(P) \le (4\mathcal{E}_\sin)^M \int \Bigg[ \prod_{A=1}^{M} \frac{d^3\vec{p}_A}{(2\pi)^3 2E(\vec{p}_A)} \Bigg] \, \delta\left( \sqrt{P^2} - \sum_{A=1}^M E(\vec{p}_A) \right) \delta^3\left( \sum_{A=1}^M \vec{p}_A \right)
    \ .
\end{gather}
We parameterize $\vec{p}_A = \lambda \vec{k}_A$ with $\lambda \ge 0$ and $\sum_{A=1}^M \| \vec{k}_A \|_2^2 = 1$. In other words, the vector $(\vec{k}_1,\dots,\vec{k}_M)$ belongs to the sphere $S^{3M-1}$. Denoting by $d\Omega_{3M-1}(\vec{k})$ the canonical integration measure on $S^{3M-1}$, we obtain
\begin{gather}
    \label{eq:cov:win-4}
    w_\sin(P) \le
    \left( \frac{\mathcal{E}_\sin}{4\pi^3} \right)^M \int_{S^{3M-1}} d\Omega_{3M-1}(\vec{k}) \, \delta^3\left( \sum_{A=1}^M \vec{k}_A \right)
    \\ \nonumber \phantom{w_\sin(P) \le } \times
    \int_0^\infty d\lambda \,
    \frac{\lambda^{3M-4}}{\prod_{A=1}^{M} E(\lambda \vec{k}_A)} \, \delta\left( \sqrt{P^2} - \sum_{A=1}^M E(\lambda \vec{k}_A) \right)
    \ .
\end{gather}
The equation $\sum_{A=1}^M E(\lambda \vec{k}_A) = \sqrt{P^2}$ admits no solution for $\sqrt{P^2} < Mm$ and a unique solution for $\sqrt{P^2} \ge Mm$, which will be denoted by $\bar{\lambda}(\sqrt{P^2},\vec{k})$. This solution is continuous and increasing in $\sqrt{P^2}$. The integral over $\lambda$ can be readily calculated, yielding
\begin{gather}
    \label{eq:cov:win-5}
    w_\sin(P) \le
    \left( \frac{\mathcal{E}_\sin}{4\pi^3} \right)^M \int_{S^{3M-1}} d\Omega_{3M-1}(\vec{k}) \, \delta^3\left( \sum_{A=1}^M \vec{k}_A \right)
    \\ \nonumber \phantom{w_\sin(P) \le } \times
    \frac{\lambda^{3M-5}}{\prod_{A=1}^{M} E(\lambda \vec{k}_A)}
    \left[ \sum_{A=1}^M \frac{ \| \vec{k}_A \|_2^2}{ E(\lambda \vec{k}_A) } \right]^{-1}_{\lambda = \bar{\lambda}(\sqrt{P^2},\vec{k})}
    \ .
\end{gather}
Using again the fact that $P$ is in $\mathbb{W}_\sin$, we derive the following inequalities
\begin{gather}
    2 \mathcal{E}_\sin \ge P_0 \ge \sqrt{P^2} = \sum_{A=1}^M E(\bar{\lambda} \vec{k}_A) \ge \sum_{A=1}^M \bar{\lambda} \| \vec{k}_A \|_2 \ge \bar{\lambda} \sum_{A=1}^M \| \vec{k}_A \|_2^2 = \bar{\lambda}
    \ , \\
    E(\bar{\lambda} \vec{k}_A) = \sqrt{ m^2 + \bar{\lambda}^2 \| \vec{k}_A \|^2 } \le m + \bar{\lambda} \| \vec{k}_A \| \le m + \bar{\lambda} \le m + 2\mathcal{E}_\sin
    \ , \\
    \sum_{A=1}^M \frac{ \| \vec{k}_A \|_2^2}{ E(\bar{\lambda} \vec{k}_A) } \ge \sum_{A=1}^M \frac{ \| \vec{k}_A \|_2^2}{ m + 2\mathcal{E}_\sin } = \frac{ 1 }{ m + 2 \mathcal{E}_\sin }
    \ ,
\end{gather}
which imply
\begin{gather}
    w_\sin(P) \le
    \frac{(2 \mathcal{E}_\sin)^{4M-5}}{(8\pi^3 m)^M} (m + 2 \mathcal{E}_\sin) \int_{S^{3M-1}} d\Omega_{3M-1}(\vec{k}) \, \delta^3\left( \sum_{A=1}^M \vec{k}_A \right)
    \nonumber \\
    \phantom{w_\sin(P)}=
    \frac{(2 \mathcal{E}_\sin)^{4M-5}}{(8\pi^3 m)^M} (m + 2 \mathcal{E}_\sin) \Omega_{3M-4}
    \ .
\end{gather}
In the last step we have noticed that the integral over $\vec{k}$ yields the surface of the sphere $S^{3M-4}$.

\section{Matrix elements in the axiomatic framework}\label{app:matrix}
We want to sketch the arguments needed to make sense and prove eq.~\eqref{eq:matrix:PsiJPsi} in the case of non-overlapping velocities. We use here the notation and symbols of section~\ref{sec:matrix}.

Translational invariance guarantees that
\begin{gather}
    \lvac \tilde{\phi}(p_{M+1}) \cdots \tilde{\phi}(p_{M+N}) \tilde{J}(q) \tilde{\phi}(p_M)^\dag \cdots \tilde{\phi}(p_1)^\dag \rvac
    \\ \nonumber
    \phantom{\lvac \tilde{\phi}(p_{M+1}) \cdots \tilde{\phi}(p_{M+N})}=
    (2\pi)^4 \delta^4 \left( q + \sum_{A=1}^{M+N} \eta_A p_A \right)
    \tilde{W}^J(p)
    \ ,
\end{gather}
for some tempered distribution $\tilde{W}^J(p)$. For every value of $t$, the matrix element
\begin{gather}
    j_t(z) = \langle \Psi_\sout(t) | J(z) | \Psi_\sin(-t) \rangle
\end{gather}
is well defined as a tempered distribution, with the explicit representation:
\begin{gather}
    \label{eq:app:matrix:jA}
    j_t(z)
    = \lvac A(f_{M+1},t) \cdots A(f_{M+N},t) J(z) A(f_M,-t)^\dag \cdots A(f_1,-t)^\dag \rvac
    \\ \nonumber =
    \int \Bigg[ \prod_{A=1}^{M+N} \frac{d^4 p_A}{(2\pi)^4} f_A^{(*)}(p_A) \Bigg]
    e^{-i \sum_{A=1}^{M+N} [ p_{A,0} {-} E(\vec{p}_A) ] t}
    e^{i \sum_{A=1}^{M+N} \eta_A p_A z}
    \tilde{W}^J(p)
    \ .
\end{gather}
Notice that, since the functions $f_A(p)$ have compact support, the above expression makes sense for any complex values of $t$ and $z$. In fact, by representing $\tilde{W}^J(p)$ as a derivative of some polynomially-bounded continuous function, one shows that $j_t(z)$ is an entire function in the variables $(t,z)$.\footnote{The fact that $j_t(z)$ is an entire function in $z$ follows also directly from the Paley-Wiener-Schwartz theorem~\cite{schwartz1950theorie}. The analogous statement for $t$ is a simple generalization of this theorem.}
In particular, the pointwise value $j_t(0)$, i.e the left-hand side of eq.~\eqref{eq:matrix:PsiJPsi}, is well defined.

Hepp~\cite{cmp/1103758732} proved that asymptotic states with Schwartz wave functions and non-overlapping velocities belong to the domain of any smeared version of $J(z)$. Therefore the matrix element
\begin{gather}
    j_\infty(z) = \langle \Psi_\sout(+\infty) | J(z) | \Psi_\sin(-\infty) \rangle
    \\ \nonumber
    \phantom{j_\infty(z)}=
    \lvac a_\sout(\check{f}_{M+1}) \cdots a_\sout(\check{f}_{M+N}) J(z) a_\sin(\check{f}_{M})^\dag \cdots a_\sin(\check{f}_{1})^\dag \rvac
\end{gather}
is well defined as a tempered distribution. Replicating the argument for $j_t(z)$, one concludes that $j_\infty(z)$ is an entire function in $z$. In particular, it can be calculated for $z=0$, which gives meaning to the right-hand side of eq.~\eqref{eq:matrix:F}.

At this point, it should be clear that both sides of eq.~\eqref{eq:matrix:PsiJPsi} are well defined, and we want to understand whether the equation itself is valid. Given a Schwartz function $g(z)$, define the smeared operator $J[g] = \int d^4z \, g(z) \, J(z)$. Assuming again non-overlapping velocities, the strong limit $\lim_{t \to +\infty} |\Psi_\as(\pm t)\rangle = |\Psi_\as(\pm \infty)\rangle$ implies
\begin{gather}
    \lim_{t \to +\infty} \langle \Psi_\sout(t) | J[g] | \Psi_\sin(-t) \rangle
    =
    \langle \Psi_\sout(+\infty) | J[g] | \Psi_\sin(-\infty) \rangle
    \ ,
\end{gather}
which is equivalently written as
\begin{gather}
    \lim_{t \to +\infty} \int d^4 z \, g(z) j_t(z) = \int d^4 z \, g(z) j_\infty(z)
    \ ,
\end{gather}
i.e. $j_t(z)$ converges to $j_\infty(z)$ in the $t \to +\infty$ limit in the weak sense. This does not automatically imply pointwise convergence of $j_t(z)$ to  $j_\infty(z)$ for $t \to +\infty$, which is what we need. Let us assume that some $C_r(R)$ exists such that the following bound holds
\begin{gather}
    \sup_{\|z\| \le R} \left| \frac{d}{dt} j_t(z) \right| \le C_r(R) t^{-r} \ ,
    \label{eq:app:matrix:bound}
\end{gather}
for any $r \ge 0$, $R > 0$ and $t > 0$. Then, the function
\begin{gather}
    \ell(z) = j_1(z) + \int_1^{+\infty} d\tau \, \frac{d}{d\tau} j_\tau(z)
\end{gather}
is well defined and
\begin{gather}
    \sup_{\|z\| \le R} \left| j_t(z) - \ell(z) \right| \le
    \int_t^{+\infty} d\tau  \sup_{\|z\| \le R} \left| \frac{d}{d\tau} j_\tau(z) \right| \le
    C_2(R) t^{-1}
    \ ,
\end{gather}
which implies that $j_t(z)$ converges uniformly to $\ell(z)$ in the $t \to +\infty$ limit on any compact subset of $\mathbb{R}^4$. If $g(z)$ is a smooth function with compact support in $\|z\| \le R$, then
\begin{gather}
    \lim_{t \to +\infty} \left| \int d^4z \, g(z) [ j_t(z) - \ell(z)] \right|
    \le
    \lim_{t \to +\infty} C_2(R) t^{-1} \int d^4z \, | g(z) |
    =
    0 \ ,
\end{gather}
i.e. $j_t(z)$ converges weakly to $\ell(z)$ in the $t \to +\infty$ limit. By the uniqueness of the weak limit, $\ell(z) = j_\infty(z)$ almost everywhere, hence everywhere since both functions are continuous. This proves that $j_t(z)$ converges uniformly to $j_\infty(z)$ in the $t \to +\infty$ limit on any compact subset of $\mathbb{R}^4$. In particular, the convergence is also pointwise.

We are left to prove the bound in eq.~\eqref{eq:app:matrix:bound}. We observe that
\begin{gather}
    \frac{d}{dt} j_t(z)
    =
    \langle \frac{d}{dt} \Psi_\sout(t) | J(z) | \Psi_\sin(-t) \rangle
    + \langle \Psi_\sout(t) | J(z) | \frac{d}{dt} \Psi_\sin(-t) \rangle
    \ .
\end{gather}
Since the analysis of the two terms is similar, we focus only on the second one. The operators $A(f,t)$ are constructed in such a way that
\begin{gather}
    \frac{d}{dt} A(f,t)^\dag \rvac = 0 \ .
\end{gather}
Using this fact, we derive
\begin{gather}
    \frac{d}{dt} | \Psi_\sin(-t) \rangle
    =
    \sum_{A=2}^M
    \Bigg\{ \lprod_{B=A+1}^{M} A(f_B,-t)^\dag \Bigg\}
    \Bigg[ \frac{d}{dt} A(f_A,-t)^\dag \, , \, \Bigg\{ \lprod_{B=1}^{A-1} A(f_B,-t)^\dag \Bigg\} \Bigg] \rvac
    \ ,
\end{gather}
where the symbol $[ \cdot , \cdot ]$ denotes the commutator. Using the definition of the operators $A(f,t)$, the definition of the state $| \Psi_\sout(t) \rangle$ and the above formula, one easily proves the following representation:
\begin{gather}
    \label{eq:app:matrix:derivative}
    \langle \Psi_\sout(t) | J(z) | \frac{d}{dt} \Psi_\sin(-t) \rangle
    =
    \sum_{A=2}^M \sum_{B=1}^{A-1}
    \int \Bigg\{ \prod_{C=1}^{M+N} d^4x_C \, \mathcal{K}_{AC}^{(*)}(x_C,t) \Bigg\}
    W^J_{AB}(x-z)
    \ ,
\end{gather}
with the definitions:
\begin{gather}
    \tilde{\zeta}_C(\tau) = \int \frac{d\omega}{2\pi} \zeta_C(\omega) e^{-i \omega \tau}
    \ , \qquad
    K_C(x) = \int \frac{d^3 \vec{p}}{(2\pi)^2}  \check{f}_C(\vec{p}) e^{-i E(\vec{p}) x_0 + i \vec{p} \vec{x}}
    \ , \\
    \mathcal{K}_{AC}(x,t) = \begin{cases}
    K_C(x) \tilde{\zeta}_C(x_{C,0}{-}\eta_C t) \quad & \text{if } A \neq C \\
    K_C(x) \frac{d}{dt} \tilde{\zeta}_C(x_{C,0}{-}\eta_C t) \quad & \text{if } A = C
    \end{cases}
    \ , \\
    W^J_{AB}(x)
    \\ \nonumber =
    \lvac \Bigg\{ \rprod_{C=M+1}^{M+N} \! \phi(x_C) \Bigg\} J(0)
    \Bigg\{ \lprod_{ \substack{B < C \le M \\ C \neq A } } \!\! \phi^\dag(x_C) \Bigg\} \,
    [ \phi^\dag(x_A) , \phi^\dag(x_B) ] \,
    \Bigg\{ \lprod_{ 1 \le C < B } \!\! \phi^\dag(x_C) \Bigg\}
    \rvac \ .
\end{gather}
Notice that $\tilde{\zeta}_C(\tau)$ is a Schwartz function and $K_C(x)$ is a smooth solution of the Klein-Gordon equation. $W^J_{AB}(x)$ is a well-defined tempered distribution (the position of $J(z)$ can be safely set to zero thanks to translational invariance), which we can represent as
\begin{gather}
    W^J_{AB}(x) = D^{\alpha} [ P_{AB}(x) F_{AB}(x) ]
\end{gather}
for some multi-index $\alpha$ (which can be chosen for convenience independent of $A$ and $B$), some polynomial $P_{AB}(x)$ and some continuous function $F_{AB}(x)$ with $\|F_{AB}\|_\infty = 1$. We introduce a smooth function $u(s)$ with the following properties: \textit{(1)} $0 \le u(s) \le 1$, \textit{(2)} $u(s) = 0$ for $s \le -1$, and \textit{(3)} $u(s) = 1$ for $s \ge -1/4$. Because of the commutator, the distribution $W^J_{AB}(x)$ vanishes if $x_A-x_B$ is spacelike, therefore
\begin{gather}
    W^J_{AB}(x-z) = u((x_A-x_B)^2) W^J_{AB}(x-z)
    \\ \nonumber =
    u((x_A-x_B)^2) D^{\alpha}_x [ P_{AB}(x-z) F_{AB}(x-z) ]
    \ .
\end{gather}
Some non-negative constants $C'$ and $s$ exist such that
\begin{gather}
| P_{AB}(x-z) | \le C' \, (1+ \|z\|_2^2)^{s} \prod_{C=1}^{M+N} (1+ \|x_C\|_2^2)^{s} \ .
\end{gather}
Using the last two equations in eq.~\eqref{eq:app:matrix:derivative}, one obtains
\begin{gather}
    \label{eq:app:matrix:derivative-2}
    \left| \langle \Psi_\sout(t) | J(z) | \frac{d}{dt} \Psi_\sin(-t) \rangle \right|
    \le
    C' \, (1+ \|z\|_2^2)^{s} \sum_{A=2}^M \sum_{B=1}^{A-1} G_{AB}(t) \prod_{\substack{ 1 \le C \le M+N \\ C \neq A,B}} G_C(t)
    \ ,
\end{gather}
with the definitions
\begin{gather}
    G_C(t) = \int d^4 x \, (1+ \|x\|_2^2)^{s} \left| D^{\alpha} \mathcal{K}_{AC}(x,t) \right|
    \ , \\
    G_{AB}(t) = \int d^4x \, d^4y \, (1+ \|x\|_2^2)^{s} (1+ \|y\|_2^2)^{s}
    \\ \nonumber \hphantom{G_{AB}(t) = \int d^4x \, d^4y \,}  \times
    \left|  D_{x}^{\alpha} D_{y}^{\alpha} \left[ \mathcal{K}_{AA}(x,t) \, \mathcal{K}_{AB}(y,t) u((x-y)^2) \right] \right|
    \ .
\end{gather}
As we will show in a moment, $G_C(t)$ is polynomially bounded and $G_{AB}(t)$ vanishes faster than any inverse power of $t$. Therefore, the product of $G_{AB}(t)$ with all $G_C(t)$ for $C \neq A,B$ vanishes faster than any inverse power of $t$. Using this fact with eq.~\eqref{eq:app:matrix:derivative-2}, one sees that, for every $r\ge 0$ a constant $C''_{r} > 0$ exists such that the following bound holds
\begin{gather}
    \left| \langle \Psi_\sout(t) | J(z) | \frac{d}{dt} \Psi_\sin(-t) \rangle \right|
    \le
    C''_r \, (1+ \|z\|_2)^{s} \, t^{-r}
\end{gather}
for any $z$ and $t>0$. An analogous bound holds for $\langle \frac{d}{dt} \Psi_\sout(t) | J(z) | \Psi_\sin(-t) \rangle$, yielding eq.~\eqref{eq:app:matrix:bound}.

We are left with the task of proving bounds for $G_C(t)$ and $G_{AB}(t)$. Since the wave functions $\check{f}_A(\vec{p})$ have compact support and the velocities $V_C$ are assumed to be non-overlapping (separately for incoming and outgoing particles), then some closed subsets $W_{C=1,\dots,M+N}$ of $\mathbb{R}^3$ exist with the following properties: \textit{(1)} $W_C$ contains a neighborhood of $V_C$, \textit{(2)} $W_C$ is a subset of the open unit ball $\{ \vec{v} \text{ s.t. } \vec{v}^2<1 \}$, \textit{(3)} the sets $W_1,\dots,W_M$ are mutually disjoint and so are the sets $W_{M+1},\dots,W_{M+N}$. With these definitions at hand, we can provide some useful bounds for the solutions of the Klein-Gordon equations $K_C(x)$. Given a four-vector $m$ of natural numbers and a real number $r \ge 0$, two constants $S_{m}$ and $T_{m,r}$ exist such that
\begin{gather}
    | D^m K_C(x) | \le
    \begin{cases}
        S_m (1 + x_0^2)^{-\frac{3}{4}} \quad & \text{for every } x
        \\[4pt]
        T_{m,r} ( 1 + x_0^2 )^{-r} (1 + \vec{x}^2 )^{-r} \quad & \text{if } \vec{x} \not\in x_0 W_C
    \end{cases}
    \ .
    \label{eq:app:matrix:Kbounds}
\end{gather}
Notice that, with no loss of generality, we have assumed $S_{m}$ and $T_{m,r}$ to be independent of $C$. In writing these inequalities we have assumed an arbitrary unit system. Units can be restored by replacing the one in the above expressions with $L^2$ where $L$ is an arbitrary length scale. The first bound in eq.~\eqref{eq:app:matrix:Kbounds} is a classical result concerning smooth solutions of the Klein-Gordon equations (in fact $D^m K_C(x)$ is a smooth solution of the Klein-Gordon equation), proved e.g. by Ruelle~\cite{Ruelle1962} (see point 1 of the lemma in page 157). The second bound in eq.~\eqref{eq:app:matrix:Kbounds} is proved e.g. by Reed and Simon~\cite{reed1979iii} (see corollary to theorem XI.14 in appendix 1 to section XI.3), Jost~\cite{jost1965general} (see \textit{Second property} in section 4, chapter VI) and also by Araki~\cite{araki1999mathematical} (see theorem 5.3 in chapter 5). Using eq.~\eqref{eq:app:matrix:Kbounds}, the inequalities
\begin{gather}
    [1+(x_0-\eta_C t)^2]^{-1} \le 4 (1+x_0^2) (1+t^2)^{-1}
    \ , \\
    1+x_0^2 \le 4 [1+(x_0-\eta_C t)^2] (1+t^2)
    \ , \\
    (1+\|x\|_2^2) \le (1+ x_0^2) (1+ \vec{x}^2)
    \ ,
\end{gather}
and the fact that $\tilde{\zeta}_C$ is Schwartz, one proves the inequalities
\begin{gather}
    \label{eq:app:matrix:Kcalbounds}
    (1+\|x\|_2^2)^s \, | D^m_x \mathcal{K}_{AC}(x,t) |
    \\ \nonumber \le
    \begin{cases}
        S'_{m,s,r} (1 + t^2)^{2s-\frac{3}{4}} \left[ 1 + (x_0 - \eta_C t)^2 \right]^{-r} \quad & \text{if } \vec{x} \in x_0 W_C
        \\[4pt]
        T'_{m,s,r} (1 + t^2)^{-r} (1+ x_0^2)^{-r} (1 + \vec{x}^2 )^{-2r} \quad & \text{if } \vec{x} \not\in x_0 W_C
    \end{cases}
    \ ,
\end{gather}
valid for any $r,s \ge 0$, any $t$ and $x$.

Using eq.~\eqref{eq:app:matrix:Kcalbounds}, one readily finds
\begin{gather}
    G_C(t)
    \le
    S'_{\alpha,s,3} (1 + t^2)^{2s-\frac{3}{4}} \int_{\vec{x} \in x_0 W_C} d^4 x \, \left[ 1 + (x_0 - \eta_C t)^2 \right]^{-3} + O(|t|^{-r})
    \\ \nonumber \le
    S'_{\alpha,s,3} (1 + t^2)^{2s-\frac{3}{4}} \text{vol}(W_C) \int d x_0 \, |x_0 + \eta_C t|^3 ( 1 + x_0^2)^{-3} + O(|t|^{-r})
    \ ,
\end{gather}
which shows explicitly that $G_C(t)$ is polynomially bounded. Let us look at $G_{AB}(t)$, first we use that the support of $u((x-y)^2)$ and all its derivatives is contained in $(x-y)^2 \ge -1$, i.e.
\begin{gather}
\label{eq:app:matrix:GAB-1}
G_{AB}(t) \le
\sum_{\beta \le \alpha} \sum_{\gamma \le \alpha} \binom{\alpha}{\beta} \binom{\alpha}{\gamma}
\int_{(x-y)^2 \ge -1} d^4x \, d^4y \, (1+ \|x\|_2^2)^{s} (1+ \|y\|_2^2)^{s}
\\ \nonumber \hphantom{G_{AB}(t) = \int d^4x \, d^4y \,}  \times
\left|
D_{x}^\beta \mathcal{K}_{AA}(x,t)
D_{y}^\gamma \mathcal{K}_{AB}(y,t)
D_x^{\alpha-\beta} D_{y}^{\alpha-\gamma} u((x-y)^2)
\right|
\ .
\end{gather}
The function $D_x^{\alpha-\beta} D_{y}^{\alpha-\gamma} u((x-y)^2)$ is bounded, i.e.
\begin{gather}
    R_{\beta+\gamma}
    =
    \sup_{x,y}
    \left|
    D_x^{\alpha-\beta} D_{y}^{\alpha-\gamma} u((x-y)^2)
    \right|
    =
    \sup_{x}
    \left|
    D_x^{\alpha-\beta-\gamma} u(x^2)
    \right|
    < + \infty
    \ .
\end{gather}
Then one can use eq.~\eqref{eq:app:matrix:Kcalbounds}, the substitutions $\vec{x} = x_0 \vec{v}$, $\vec{y} = y_0 \vec{w}$, and subsequently the substitutions $x_0 \to x_0-t$, $y_0 \to y_0-t$ to derive
\begin{gather}
\label{eq:app:matrix:GAB-2}
G_{AB}(t) \le
(1 + t^2)^{4s-\frac{3}{2}}
\sum_{\beta \le \alpha} \sum_{\gamma \le \alpha} \binom{\alpha}{\beta} \binom{\alpha}{\gamma} S'_{\beta,s,r} S'_{\gamma,s,r} R_{\beta+\gamma}
\int_{W_A} d^3 \vec{v} \int_{W_B} d^3 \vec{w}
\\ \nonumber \hphantom{G_{AB}(t) \le}  \times
\int_{ \substack{ [x_0\vec{v} -y_0\vec{w} - t(\vec{v}-\vec{w})]^2 \\ \le 1 + (x_0-y_0)^2 } } dx_0 \, dy_0 \,
\frac{ |x_0-t|^3 \, |y_0-t|^3 }{ ( 1 + x_0^2 )^r ( 1 + y_0^2 )^r }
+ O(t^{-r})
\ .
\end{gather}
Notice that, assuming $t>0$, if $(x_0,y_0)$ belongs to the integration domain then
\begin{gather}
    1 + |x_0| + |y_0| \ge 1 + |x_0-y_0| \ge \sqrt{1 + (x_0-y_0)^2}
    \\ \nonumber
    \ge \left| x_0\vec{v} -y_0\vec{w} - t(\vec{v}-\vec{w}) \right|
    \ge t | \vec{v} - \vec{w} | - |x_0| \, |\vec{v}|  - |y_0| \, |\vec{w}|
    \ge t d_{AB} - |x_0|  - |y_0|
    \ ,
\end{gather}
where $d_{AB}>0$ is the distance between the sets $W_A$ and $W_B$. Therefore, if $(x_0,y_0)$ belongs to the integration domain and $t \ge 2/d_{AB}$, the following inequality holds
\begin{gather}
    \max\{|x_0|,|y_0|\} \ge \frac{|x_0| + |y_0|}{2} \ge \frac{t d_{AB} - 1}{4} \ge \frac{t d_{AB}}{8} \ .
\end{gather}
This inequality can be used to enlarge the integration domain in eq.~\eqref{eq:app:matrix:GAB-2}, yielding
\begin{gather}
\label{eq:app:matrix:GAB-3}
G_{AB}(t) \le
2 (1 + t^2)^{4s-\frac{3}{2}} \text{vol}(W_A) \, \text{vol}(W_B)
\sum_{\beta \le \alpha} \sum_{\gamma \le \alpha} \binom{\alpha}{\beta} \binom{\alpha}{\gamma} S'_{\beta,s,r} S'_{\gamma,s,r} R_{\beta+\gamma}
\\ \nonumber \hphantom{G_{AB}(t) \le}  \times
t^4
\int dy_0 \,
\frac{ |y_0-t|^3 }{ ( 1 + y_0^2 )^r }
\int_{ |\sigma| \ge \frac{d_{AB}}{8} } d\sigma \,
\frac{ |\sigma-1|^3 }{ ( 1 + t^2 \sigma^2 )^r }
+ O(t^{-r})
\ .
\end{gather}
Since $r$ is an arbitrary positive constant, we conclude that $G_{AB}(t)$ vanishes faster than any inverse power of $t$ for $t \to +\infty$.

\clearpage
\small
\addcontentsline{toc}{section}{References}
\bibliographystyle{JHEP}
\bibliography{non-inspire,inspire}

\end{document}